\documentclass[twocolumn,aps,prl]{revtex4-1}

\usepackage[usenames,dvipsnames]{xcolor}
\usepackage{amsmath}
\usepackage{amstext}
\usepackage{amsfonts}
\usepackage{amssymb}
\usepackage[colorlinks=true,citecolor=blue,linkcolor=red]{hyperref}
\usepackage{graphicx}
\usepackage{bbold}					
\usepackage[makeroom]{cancel}		
\usepackage{multirow}				
\usepackage[normalem]{ulem}        
\usepackage{array}
\usepackage{pdfpages}
\makeatletter
\AtBeginDocument{\let\LS@rot\@undefined}
\makeatother

\newcolumntype{x}[1]{>{\centering\let\newline\\\arraybackslash\hspace{0pt}}p{#1}}

\renewcommand{\Im}{\operatorname{Im}}

\DeclareMathAlphabet{\mathbbold}{U}{bbold}{m}{n}

\newcounter{subeqn} %
\makeatletter
\@addtoreset{subeqn}{equation}
\makeatother

\definecolor{TB}{rgb}{0,0,0} 

\begin{document}

\title{Non-Hermitian skin modes induced by on-site dissipations and chiral tunneling effect}

\author{Yifei Yi$^{1,2}$}
\author{Zhesen Yang$^{1,2}$}\email[Corresponding author: ]{yangzs@iphy.ac.cn}

\affiliation{$^{1}$Beijing National Laboratory for Condensed Matter Physics,
	and Institute of Physics, Chinese Academy of Sciences, Beijing 100190, China}
\affiliation{$^{2}$University of Chinese Academy of Sciences, Beijing 100049, China}

\date{\today}
	
\begin{abstract}
In this paper, we study the conditions under which on-site dissipations can induce non-Hermitian skin modes in non-Hermitian  systems. When the original Hermitian Hamiltonian has spinless time-reversal symmetry, it is impossible to have skin modes; on the other hand, if the Hermitian Hamiltonian has spinful time-reversal symmetry, skin modes can be induced by on-site dissipations under certain circumstance. As a concrete example, we employ the Rice-Mele model to illustrate our results. Furthermore, we predict that the skin modes can be detected by the chiral tunneling effect, that is, the tunneling favors the direction where the skin modes are localized. Our work reveals a no-go theorem for the emergence of skin modes, and paves the way for searching for quantum systems with skin modes and studying their novel physical responses.
\end{abstract}
	
\maketitle

{\em Introduction.}---Non-Hermitian Hamiltonians~\cite{benderMakingSenseNonHermitian2007b,Rotter_2009,moiseyevNonHermitianQuantumMechanics2011}, which describe the nonconservative phenomena~\cite{miriExceptionalPointsOptics2019a}, have been widely studied  recently~\cite{miriExceptionalPointsOptics2019a,fengNonHermitianPhotonicsBased2017b,el-ganainyNonHermitianPhysicsPT2018d,ozdemirParityTimeSymmetry2019a,RevModPhys.91.015006,PhysRevLett.118.045701,PhysRevLett.121.203001,PhysRevLett.123.123601,2019arXiv190912516P,2019arXiv191003229L,martinezalvarezTopologicalStatesNonHermitian2018b,Foa_Torres_2019,ghatakNewTopologicalInvariants2019,2019arXiv191210048B,PhysRevLett.77.570,PhysRevLett.102.065703,2017arXiv170805841K,PhysRevB.99.201107,PhysRevLett.121.026403,PhysRevB.98.035141,PhysRevB.98.245130,PhysRevB.99.041116,PhysRevLett.123.097701,2019arXiv191205825L,2019arXiv191212022Y,2020arXiv200100697L,PhysRevLett.124.040401,PhysRevLett.123.170401}. It has been shown that some non-Hermitian Hamiltonians with open boundary condition can never be characterized by Bloch Hamiltonians~\cite{yaoEdgeStatesTopological2018b,yaoNonHermitianChernBands2018b,PhysRevLett.123.170401,PhysRevLett.123.246801,yokomizoNonBlochBandTheory2019a,zhangCorrespondenceWindingNumbers2019,2019arXiv191205499Y,okumaTopologicalOriginNonHermitian2019,xiongWhyDoesBulk2018c,kunstBiorthogonalBulkBoundaryCorrespondence2018a,martinezalvarezNonHermitianRobustEdge2018b,leeAnatomySkinModes2019b,2019arXiv191206974L,longhiProbingNonHermitianSkin2019,PhysRevB.99.081103,PhysRevLett.122.076801,PhysRevLett.123.016805,herviouDefiningBulkedgeCorrespondence2019,PhysRevB.99.155431,PhysRevB.100.054301,zirnsteinBulkboundaryCorrespondenceNonHermitian2019a,PhysRevLett.124.056802,PhysRevB.100.045141, dengNonBlochTopologicalInvariants2019a,PhysRevLett.123.073601,brzezickiHiddenChernNumber2019,PhysRevB.100.165430,longhiNonBlochPTSymmetry2019,2019arXiv191204024W,	2019arXiv191207687K}. To be more precise, the open boundary spectra may collapse compared to the periodic boundary spectra, along with the emergence of non-Hermitian skin modes~\cite{yaoEdgeStatesTopological2018b}. It has been shown that both phenomena can be well understood with the concept of generalized Brillouin zone (GBZ)~\cite{yaoEdgeStatesTopological2018b,yaoNonHermitianChernBands2018b,PhysRevLett.123.246801,yokomizoNonBlochBandTheory2019a,zhangCorrespondenceWindingNumbers2019,2019arXiv191205499Y}, which is a generalization of Brillouin zone (BZ) defined in systems (Hermitian or non-Hermitian) with open boundaries. When the GBZ coincides with the BZ, the open boundary spectra can be described by the Bloch Hamiltonian with no skin modes, and the conventional bulk-boundary correspondence still holds. On the other hand, if GBZ is distinct from BZ, the open boundary spectra collapses, and skin modes along with the anomalous bulk-boundary correspondence emerge at the same time~\cite{yaoEdgeStatesTopological2018b}. Inspired by the theoretical proposal, non-Hermitian skin modes have been observed experimentally in the classical wave systems recently~\cite{2019arXiv190711619G,2019arXiv190711562H,1907.12566,2019arXiv190802759H}. Finding the conditions for the emergence of skin modes in quantum systems and investigating the corresponding novel physical responses are interesting and challenging~\cite{PhysRevB.84.205128,leeAnomalousEdgeState2016a,leykamEdgeModesDegeneracies2017e,PhysRevLett.120.146402,PhysRevB.97.045106,PhysRevA.97.052115,PhysRevX.8.031079,Kawabata:2019aa,PhysRevX.9.041015,PhysRevB.98.115135,PhysRevB.99.125103,PhysRevB.99.235112,PhysRevLett.123.206404,Zhou1009,Cerjan:2019aa,PhysRevLett.123.237202,PhysRevA.98.042114,PhysRevB.99.081102,PhysRevB.99.041406,PhysRevB.99.121101,PhysRevLett.123.066405,2019arXiv190500210Y,2019arXiv191202788Y,2019arXiv190504338S,2019arXiv191112748W,PhysRevLett.122.237601,PhysRevLett.123.183601,PhysRevLett.123.190403,PhysRevA.98.052116,PhysRevB.99.245116,PhysRevLett.124.046401,PhysRevB.100.075403,2019arXiv190603988H,PhysRevLett.124.066602}.

On-site dissipations are well-controlled non-Hermitian terms that can be realized experimentally in both classical and quantum systems~\cite{miriExceptionalPointsOptics2019a,fengNonHermitianPhotonicsBased2017b,el-ganainyNonHermitianPhysicsPT2018d,ozdemirParityTimeSymmetry2019a,RevModPhys.91.015006,Bouganne:2020aa,Tomitae1701513,PhysRevA.99.031601,Li:2019aa,Syassen1329,PhysRevLett.110.035302,PhysRevResearch.2.022051}. In contrast to the non-reciprocal terms, like $\sum_i (t\hat{a}_i^\dag\hat{b}_i+2t\hat{b}_i^\dag\hat{a}_i)$, on-site dissipations, such as $\sum_i(i\gamma_a\hat{a}_i^\dag\hat{a}_i+i\gamma_b\hat{b}_i^\dag\hat{b}_i)$, do not favor any special hopping direction. Although it has been revealed that skin modes can be induced by on-site dissipations~\cite{PhysRevLett.123.170401,1907.12566,2019arXiv191003229L,martinezalvarezNonHermitianRobustEdge2018b}, their exact relation is still unclear.

In this paper, we show that if a Hermitian non-superconducting system has spinless time-reversal symmetry (TRS), on-site dissipations will not induce non-Hermitian skin modes. However, if the Hermitian system has spinful TRS, it is possible for the system to have skin modes, depending on whether the system has inversion symmetry (IS) and its representation. As a concrete example, we use Rice-Mele model~\cite{PhysRevLett.49.1455} to illustrate our results. The novel physical responses of skin modes are also investigated.

\renewcommand\arraystretch{1.5}
\begin{table*}[t]
	\caption{\label{T1}
	Non-Hermitian symmetry ramifications of Eq.~\ref{effH}. All the elements of 1D Hermitian (non-Hermitian) symmetry groups are listed in the first (second and fifth) row. If $\mathcal{H}_s(k)$ has one of the eight Hermitian symmetries listed in the first row, then depending on the commutation relation, the corresponding non-Hermitian symmetries, listed in the second and fifth rows, will be preserved for $\mathcal{H}_{s,eff}(k)$. The third and sixth rows represent the symmetry constraints to the characteristic equation $f(\beta,E)=\det[E-\mathcal{H}(\beta)]$, where $\mathcal{H}(\beta)$ is the non-Bloch Hamiltonian with $\beta=e^{ik}$ and $k\in\mathbb{C}$. In the SM \uppercase\expandafter{\romannumeral3} D, we use an example to illustrate the application of Table.~\ref{T1}.}
	\label{t1}
	\begin{tabular*}{17.86cm}{|p{1.9cm}|p{2.2cm}|p{1.5cm}|p{1.5cm}|p{1.6cm}|p{1.5cm}|p{1.5cm}|p{1.5cm}|p{1.6cm}|p{1.5cm}|}
		\hline
		& Hermitian                                & $\rm{I}$                                & $\mathcal{PT}$       & $\mathcal{P}$                             & $\mathcal{T}$       & $\mathcal{TC}$                          & $\mathcal{PC}$       & $\mathcal{PTC}$                           & $\mathcal{C}$       \\ \hline
		\multirow{3}{*}{$[\Gamma_0,U_X]=0$}   & Non-Hermitian                         & $\rm{I}$                                & $\mathcal{P\bar{T}}$ & $\mathcal{P}$                             & $\mathcal{\bar{T}}$ & $\mathcal{TC},\mathcal{\bar{T}\bar{C}}$ & $\mathcal{P\bar{C}}$ & $\mathcal{PTC},\mathcal{P\bar{T}\bar{C}}$ & $\mathcal{\bar{C}}$ \\ \cline{2-10}
		& $U_X^{-1}\mathcal{H}(k)U_X=$             & $\mathcal{H}(k)$~                        & $\mathcal{H}^t(k)$~   & $\mathcal{H}(-k)$                         & $\mathcal{H}^t(-k)$~ & $-\mathcal{H}^\dag(k)$~                       & $-\mathcal{H}^*(k)$~  & $-\mathcal{H}^\dag(-k)$                   & $-\mathcal{H}^*(-k)$~ \\ \cline{2-10}
		& $f(\beta,E)=$ & \multicolumn{2}{c|}{$f(\beta,E)$}                              & \multicolumn{2}{c|}{$f(1/\beta,E)$}                             & \multicolumn{2}{c|}{$f(1/\beta^*,-E^*)$}                       & \multicolumn{2}{c|}{$f(\beta^*,-E^*)$}                          \\ \hline
		\multirow{3}{*}{$\{\Gamma_0,U_X\}=0$} & Non-Hermitian                            & $\mathcal{T\bar{T}},\mathcal{C\bar{C}}$ & $\mathcal{PT}$       & $\mathcal{PT\bar{T}},\mathcal{PC\bar{C}}$ & $\mathcal{T}$       & $\mathcal{\bar{T}C},\mathcal{T\bar{C}}$ & $\mathcal{PC}$       & $\mathcal{P\bar{T}C},\mathcal{PT\bar{C}}$ & $\mathcal{C}$       \\ \cline{2-10}
		& $U_X^{-1}\mathcal{H}(k)U_X=$             & $\mathcal{H}^\dag(k)$                   & $\mathcal{H}^*(k)$   & $\mathcal{H}^\dag(-k)$                    & $\mathcal{H}^*(-k)$ & $-\mathcal{H}(k)$                  & $-\mathcal{H}^t(k)$  & $-\mathcal{H}(-k)$                        & $-\mathcal{H}^t(-k)$ \\ \cline{2-10}
		& $f(\beta,E)=$ & \multicolumn{2}{c|}{$f(1/\beta^*,E^*)$}                        & \multicolumn{2}{c|}{$f(\beta^*,E^*)$}                           & \multicolumn{2}{c|}{$f(\beta,-E)$}                             & \multicolumn{2}{c|}{$f(1/\beta,-E)$}                            \\ \hline
	\end{tabular*}
\end{table*}

{\em Non-Hermitian Hamiltonians with on-site dissipations.}---We start from the following one-dimensional (1D) {\em Hermitian} Hamiltonian,
\begin{equation}
\hat{H}=\hat{H}_s+\hat{H}_{b}+\hat{H}_{s-b}.
\label{Hamiltonian}
\end{equation}
Here $\hat{H}_s=\sum_{i,j}\sum_{\mu,\nu}t_{ij}^{\mu\nu}\hat{c}_{i\mu}^\dag\hat{c}_{j\nu}$ is the system Hamiltonian we concerned, where $i,j$ and $\mu,\nu$ label lattice sites and band (or spin) indexes, respectively; $\hat{H}_{b}=\sum_{i, p_\mu,\mu}\left(\varepsilon_{ p_\mu}-\mu_{p_\mu}\right) \hat{b}_{i p_\mu}^{\dagger} \hat{b}_{i p_\mu}$ comes from a free Fermion bath, where $p_\mu$ is the internal degrees of the bath; and $\hat{H}_{s-b}=\sum_{i, p_\mu,\mu} V_{p_\mu\mu}(\hat{c}_{i\mu}^{\dagger} \hat{b}_{i p_\mu}+\hat{b}_{i p_\mu}^{\dagger} \hat{c}_{i\mu})$ is the system-bath coupling term. We first focus on the periodic boundary condition. In the Supplemental Materials \uppercase\expandafter{\romannumeral1} (SM \uppercase\expandafter{\romannumeral1}), we show that the following non-Hermitian effective Hamiltonian can be obtained by using the standard Green's function method~\cite{RevModPhys.86.779},
\begin{equation}
\mathcal{H}_{s,eff}(k)=\mathcal{H}_s(k)-i\gamma\Gamma_0,\quad \mathcal{H}_s(k)=\mathcal{H}^\dag_s(k),
\label{effH}
\end{equation}
where $[\mathcal{H}_s(k)]_{\mu \nu}=\sum_{l_{\mu\nu}} t_{l_{\mu\nu}}^{\mu \nu} e^{ikl_{\mu\nu}}$ is the Bloch Hamiltonian of the system; $\gamma$ is proportional to the density of states (DoS) of the external bath and the system-bath coupling strength; and $\Gamma_0$ is a diagonal matrix, whose matrix elements represent the dissipations for each band (or spin). This kind of dissipation is dubbed as {\em on-site dissipation}
in this paper. Exploring the condition for the emergence of skin modes in Eq.~\ref{effH} is the central topic of this paper. The extension to the general non-Hermitian Hamiltonians will be discussed in the final section.

{\em Non-Hermitian symmetries and skin modes.}---The main results of this paper can be summarized as follows. If $\mathcal{H}_s(k)$ in Eq.~\ref{effH} preserves TRS but breaks particle-hole symmetry (PHS)~\cite{SM3}, then, (i) for the spinless case, it is impossible to have skin modes; (ii) for the spinful case, if the skin modes are to emerge, one of the following three conditions must be satisfied: (a) $\mathcal H_s(k)$ breaks IS; (b) $\mathcal H_s(k)$ preserves IS represented by $\mathcal{P}$, but $\{\mathcal{P},\Gamma_0\}=0$; (c) $\mathcal H_s(k)$ preserves IS and $[\mathcal{P},\Gamma_0]=0$, but $\{\mathcal{P},\mathcal{T}\}=[\Gamma_0,\mathcal{T}]=0$, where $\mathcal{T}$ represents TRS. While if $\Gamma_0$ and the symmetries of $\mathcal H_s(k)$ do not satisfy the above three conditions, it is impossible to have skin modes. Our results reveal a no-go theorem for the emergence of skin modes. For example, if $\mathcal{T}=i\sigma_y\mathcal{K}^*$, where $\mathcal{K}^*$ represents the complex conjugate operator, $\mathcal{P}=\tau_z$, it is impossible to have skin modes in Eq.~\ref{effH}~\cite{SM4}. However, if $\mathcal{T}=i\sigma_y\mathcal{K}^*$, $\mathcal{P}=\tau_x$, skin modes can be induced by the on-site dissipation $\Gamma_0=\tau_z$ due to $\{\mathcal{P},\Gamma_0\}=0$.

Our derivation of the main results is based on the GBZ theory~\cite{yaoEdgeStatesTopological2018b,yokomizoNonBlochBandTheory2019a,zhangCorrespondenceWindingNumbers2019,2019arXiv191205499Y}. Here we briefly summarize the procedure of our derivation. All the details can be found in SM~\cite{SM1}. We first write down all the non-Hermitian symmetry groups that Eq.~\ref{effH} belongs to when $\mathcal{H}_s(k)$ preserves TRS but breaks PHS. After that, we use the GBZ theory to derive which non-Hermitian symmetry groups forbid the emergence of skin modes. As shown in Table.~\ref{T1}, all the elements of 1D Hermitian (non-Hermitian) symmetry groups are listed in the first (second and fifth) row. Here $\mathcal{T,C,P,\bar{T},\bar{C}}$ represent TRS, PHS, IS, anomalous time-reversal symmetry (TRS$^{\dag}$), and anomalous particle-hole symmetry (PHS$^{\dag}$), respectively~\cite{PhysRevX.8.031079,PhysRevX.9.041015}, while the others represent the combination of the above five symmetries, e.g., $\mathcal{PT}$ represents the combination of TRS and IS~\cite{SM5}. The symmetry constraints to the Bloch Hamiltonian are summarized in the third and sixth rows of Table.~\ref{T1}~\cite{PhysRevX.9.041015}. We note that the derivation of skin modes requires the information of the GBZ Hamiltonian, which is an extension of the Bloch Hamiltonian to the entire complex plane via a substitution, $\mathcal{H}(k) \rightarrow\mathcal{H}\left(\beta=e^{i k}\right)$ where $k\in\mathbb{C}$~\cite{yaoEdgeStatesTopological2018b,yokomizoNonBlochBandTheory2019a,2019arXiv191205499Y}. In the SM \uppercase\expandafter{\romannumeral2}, we show how symmetries constrain the characteristic equation, $f(\beta,E)=\det[E-\mathcal{H}(\beta)]$, and the result can be summarized in the fourth and seventh rows of Table.~\ref{T1}. 

In Eq.~\ref{effH}, an important observation is that all the non-Hermitian symmetries of $\mathcal{H}_{s,eff}(k)$ have a Hermitian origin. For example, when $\mathcal{H}_s(k)$ preserves TRS represented by $U_\mathcal{T}\mathcal{K}^*$, it automatically preserves TRS$^{\dag}$ due to $\mathcal{H}_s(k)=\mathcal{H}_s^\dag(k)$. It can be deduced that if $[\Gamma_0,U_{\mathcal{T}}]=0$, then, TRS is broken but TRS$^\dag$ is preserved for the overall non-Hermitian Hamiltonian $\mathcal{H}_{s,eff}(k)$. On the other hand, if $\{\Gamma_0,U_{\mathcal{T}}\}=0$, TRS$^\dag$ is broken but TRS is preserved. This phenomena is called symmetry ramification~\cite{PhysRevX.9.041015}. In the SM \uppercase\expandafter{\romannumeral3}, we show that the ramification for other symmetries obeys a similar rule as shown in Table.~\ref{T1}. Finally, if $\mathcal{H}_s(k)$ preserves TRS but breaks PHS, in the SM \uppercase\expandafter{\romannumeral3}, we show that $\mathcal{H}_{s,eff}(k)$ belongs to the following non-Hermitian symmetry groups, $G_{\mathcal{T}_{\pm}}$,  $G_{\mathcal{T}_{\pm},(\mathcal{PT})_{\pm}}$,
$G_{\mathcal{T}_{\pm},(\mathcal{PC})_{\pm}}$,
$G_{\mathcal{T}_{\pm},(\mathcal{P\bar{T}})_{\pm}}$,
$G_{\mathcal{T}_{\pm},(\mathcal{P\bar{C}})_{\pm}}$,
$G_{\mathcal{\bar{T}}_{\pm}}$,
$G_{\bar{\mathcal{T}}_{\pm},(\mathcal{PT})_{\pm}}$,  $G_{\mathcal{\bar{T}}_{\pm},(\mathcal{PC})_{\pm}}$,
$G_{\bar{\mathcal{T}}_{\pm},(\mathcal{P\bar{T}})_{\pm}}$,  $G_{\mathcal{\bar{T}}_{\pm},(\mathcal{P\bar{C}})_{\pm}}$. Here $G$ represents the group generators. For example,  $G_{\mathcal{T}_{-},(\mathcal{PT})_{+}}=\{{\rm{I}} \text{ (identity element)},\mathcal{P}=U_{\mathcal{P}},\mathcal{T}_{-}=U_{\mathcal{T}_{-}}\mathcal{K}^*,(\mathcal{PT})_{+}=U_{\mathcal{(PT)}_{+}}\mathcal{K}^*\}$, with $U_{\mathcal{T}_{-}}U_{\mathcal{T}_{-}}^*=-1$ and $U_{\mathcal{(PT)}_{+}}U_{\mathcal{(PT)}_{+}}^*=1$.

For {\em all} the symmetry groups listed above, skin modes are absent when $G$ contains spinless TRS$^\dag$ ($\bar{\mathcal{T}}_+$) or IS ($\mathcal{P}$) or both (see SM \uppercase\expandafter{\romannumeral4} for details). An exceptional case is $G_{\bar{\mathcal{T}}_-,(\mathcal{P\bar{T}})_+}$, in which skin modes emerge with the presence of IS. Note that this result can be generalised to apply to non-Hermitian Hamiltonians of any form. The derivation is based on the constraints that symmetries impose on the characteristic equation (shown in the fourth and seventh rows of Table.~\ref{T1})~\cite{SM1}, and the GBZ condition (shown in the SM \uppercase\expandafter{\romannumeral4})~\cite{yaoEdgeStatesTopological2018b,yokomizoNonBlochBandTheory2019a,zhangCorrespondenceWindingNumbers2019,2019arXiv191205499Y,new,GBZ_intro}. For example, if the non-Hermitian system has and only has TRS$^\dag$, according to the fourth row of Table.~\ref{T1}, the characteristic equation satisfies $f(\beta,E)=f(1/\beta,E)$. The GBZ conditions for the systems with spinless ($\mathcal{\bar{T}}_+$) case and spinful ($\mathcal{\bar{T}}_-$) case  are $|\beta_p(E)|=|\beta_{p+1}(E)|$ and  $|\beta_{p-1}(E)|=|\beta_p(E)|\ \&\ |\beta_{p+1}(E)|=|\beta_{p+2}(E)|$, respectively, where $\beta_i$ is the $i$th largest root (order by absolute value) of $f(\beta,E)=0$, and $p$ is the order of the pole of $f(\beta,E)=0$. Therefore, $\mathcal{\bar{T}}_+$ forbids the emergence of skin modes~\cite{SM6}, while $\mathcal{\bar{T}}_-$ does not~\cite{SM7}. In SM \uppercase\expandafter{\romannumeral4}, we also provided numerical verifications for all the symmetry groups we concerned, which are consistent with our derivations.

\begin{figure}[t]
	\centerline{\includegraphics[height=6.1cm]{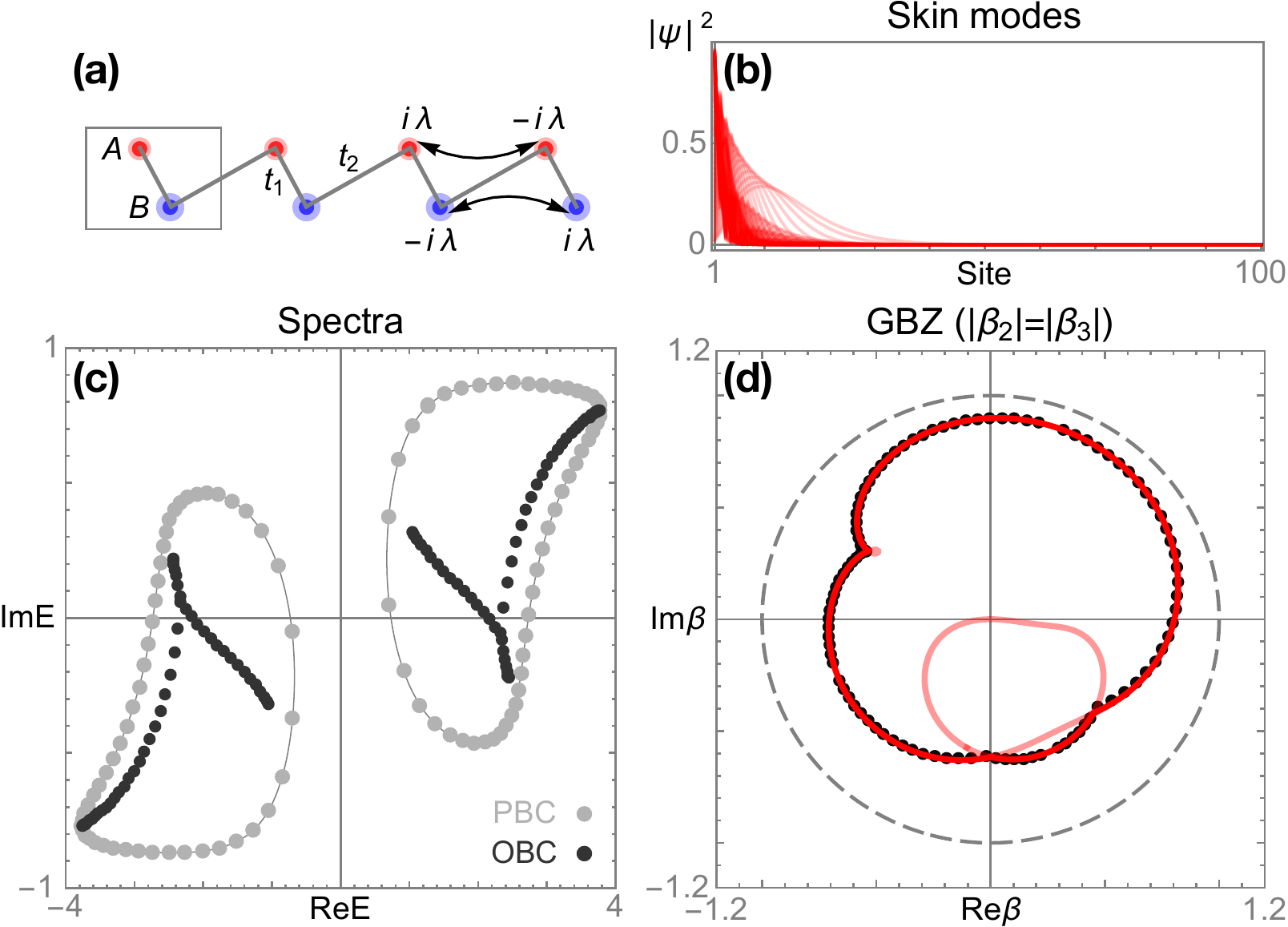}}
	\caption{Skin modes induced by the on-site dissipation in the TRS breaking Rice-Mele model, i.e., Eq.~\ref{spinless}. (a) shows the schematic diagram of the {\em Hermitian} part, namely, $\mathcal{H}_{\rm RM}(k)+\lambda\sin k\sigma_z$. (b), (c), and (d) show all the eigenstates (skin modes),  open/periodic boundary condition spectrum,  and numerical result of GBZ (black points) and auxiliary GBZ~\cite{2019arXiv191205499Y} (red lines) of the system, respectively.
		\label{F1}}
\end{figure}

Now we show that the conclusions in the last paragraph are equivalent to the main results discussed at the beginning of this section. When $\mathcal{H}_s(k)$ has spinless TRS with $\mathcal{T}_+=\mathcal{K}^*$, $\mathcal{H}_{s,eff}(k)$ must preserve spinless TRS$^\dag$ with $\mathcal{\bar{T}}_+=\mathcal{K}^t$ due to $[U_{\mathcal{T}_+},\Gamma_0]=0$. Thus it is impossible to have skin modes. This is the main result (i) discussed above. For the spinful case with $\mathcal{T}_-=U_{\mathcal{T}_-}\mathcal{K}^*$, if $\mathcal{H}_s(k)$ has $\mathcal{P}=U_{\mathcal{P}}$ and $[U_\mathcal{P},\Gamma_0]=0$, the non-Hermitian Hamiltonian must preserve IS. This forbids the emergence of skin modes in general. For the exceptional case, the existence of $\mathcal{\bar{T}}_-$ and $(\mathcal{P\bar{T}})_+$ symmetries implies the Hermitian Hamiltonian $\mathcal{H}_s(k)$ must preserve $\mathcal{T}_-$ and $(\mathcal{PT})_+$ symmetries, which ultimately leads to $\{\mathcal{P},\mathcal{T}\}=0$~\cite{RevModPhys.88.035005,PhysRevB.88.075142,PhysRevB.90.205136,SM1}, and this is equivalent to the main result (ii).

{\em Example.}---In order to verify our results, we use Rice-Mele model as an example
\begin{equation}
\mathcal{H}_{\rm{RM}}(k)=(t_1+t_2\cos k)\sigma_x+t_2\sin k\sigma_y+\mu\sigma_z,
\label{RMm}
\end{equation}
which preserves $\mathcal{T}_+=\mathcal{K}^*, (\mathcal{PC})_-=\sigma_y\mathcal{K}^t, \mathcal{PCT}=\sigma_y\mathcal{K}^\dag$. Since Eq.~\ref{RMm} preserves $\mathcal{T}_{+}$ and breaks $\mathcal{P}$, in order to induce skin modes with on-site dissipations, we can either break TRS or add spin-orbit coupling (see SM \uppercase\expandafter{\romannumeral5} for details). For the spinless case, as shown in Fig.~\ref{F1} (a), we study the case where Rice-Mele model breaks TRS,
\begin{equation}
\mathcal{H}_{\rm{spinless}}(k)=\mathcal{H}_{\rm{RM}}(k)+\lambda\sin k\sigma_z+i\gamma\sigma_z,
\label{spinless}
\end{equation}
where $\lambda$ controls the term that breaks TRS. It is easy to verify that only $(\mathcal{PC})_-$ symmetry is preserved for Eq.~\ref{spinless}, which implies $f(\beta,E)=f(\beta,-E)$. According to the GBZ condition $|\beta_p|=|\beta_{p+1}|$ shown in SM \uppercase\expandafter{\romannumeral4}, we can deduce that (i) the spectrum is formed by pairs as $(E,-E)$; (ii) the roots of the characteristic equation satisfy  $\beta(E)=\beta(-E)$, which means the sub-GBZs~\cite{2019arXiv191205499Y} for the $E$ and $-E$ bands are the same. All the wavefunctions of Eq.~\ref{spinless} with $t_1=\lambda=2,t_2=\mu=\gamma=1$ and $N=100$ (lattice site) are plotted in Fig.~\ref{F1} (b), one can notice that they all localize at the left boundary. The discrepancy between periodic and open boundary spectrum, as shown in Fig.~\ref{F1} (c)~\footnote{We note in Fig.~\ref{F1} (c)-(d), the lattice size $N$ is selected as $50$, while in Fig.~\ref{F1} (b), $N$ is selected as $100$}, also reveals the existence of skin modes~\cite{yaoEdgeStatesTopological2018b,zhangCorrespondenceWindingNumbers2019,kunstNonHermitianSystemsTopology2019b,2018arXiv181202011L}. The corresponding numerical calculation of GBZ (black points) and auxiliary GBZ~\cite{2019arXiv191205499Y} (red lines) are shown in Fig.~\ref{F1} (d), which are both inside the unit circle (gray dashed lines). In the SM  \uppercase\expandafter{\romannumeral5}, we show that regardless the value of $\mu$, the skin modes exist when $\lambda t_1t_2\neq0$. This means the on-site dissipations can also induce skin modes in the Su-Schrieffer-Heeger model~\cite{PhysRevLett.42.1698} when TRS is broken.

\begin{figure}[b]
	\centerline{\includegraphics[height=4.35cm]{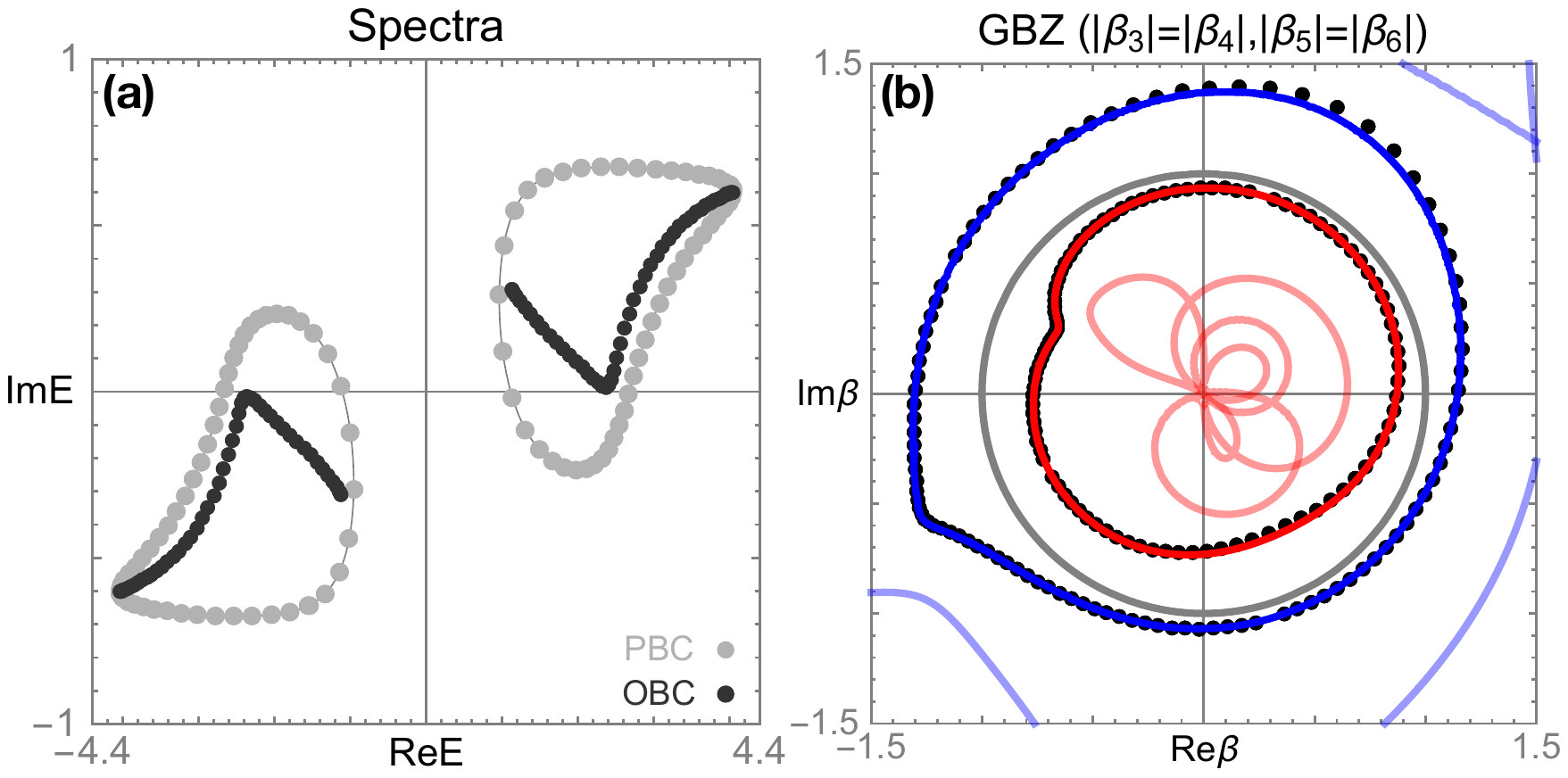}}
	\caption{$Z_2$ skin modes induced by the on-site dissipation in the Rice-Mele model with spin-orbit coupling, i.e., Eq.~\ref{spinful}. (a) and (b) show the periodic/open boundary condition spectrum and the corresponding auxiliary GBZ (solid lines) and numerical calculated GBZ (Block points) of the system, respectively. Notice that the GBZ condition for the system with spinful TRS$^\dag$ is $|\beta_{p-1}|=|\beta_p|\ \&\ |\beta_{p+1}|=|\beta_{p+2}|$, where $p=4$ in our model.
		\label{F2}}
\end{figure}

For the spinful case, since the Rice-Mele model breaks IS, on-site dissipation can induce skin modes if we add spin-orbit coupling. Therefore, the following Bloch Hamiltonian with intrinsic and shortest ranged Rashiba spin-orbit coupling~\cite{PhysRevB.96.235130} is studied
\begin{equation}\begin{aligned}
&\mathcal{H}_{{\rm spinful}}(k)=\mathcal{H}_{{\rm RM}}(k)s_0+\mathcal{H}_{{\rm soc}}(k)+i\gamma\sigma_zs_0,\\
&\mathcal{H}_{{\rm soc}}(k)=\lambda_I \sin k\sigma_zs_z-\lambda_R\sigma_y(s_x-\sqrt{3}s_y)/2,
\end{aligned}\label{spinful}
\end{equation}
where $s$ is the spin Pauli matrix. Under the action of spinful TRS$^\dag$, $|\beta,E,\uparrow\rangle$ maps to $|1/\beta,E,\downarrow\rangle$. Therefore, a left localized eigenstate with $|\beta|<1$ will be mapped to the right one with  $|\beta|>1$. These skin modes are called $Z_2$ skin modes~\cite{okumaTopologicalOriginNonHermitian2019} and protected by TRS$^\dag$. Indeed, according to the GBZ condition, we require $|\beta_{p-1}|=|\beta_p|=1/r_0$ for one spin band, and $|\beta_{p+1}|=|\beta_{p+2}|=r_0$ for the other. The absence of IS implies there is no guarantee for $1/r_0=r_0$. Therefore, skin modes can emerge. This can be checked by the comparison of open/periodic boundary spectrum and the corresponding GBZ shown in Fig.~\ref{F2} with the following parameters $t_1=\lambda_I=2,t_2=\mu=\lambda_R=\gamma=1,N=50$. As shown in (b), the GBZ for one spin band (the red lines containing the black points) is larger than 1, and the other (the blue lines containing the black points) is smaller than 1. In the SM  \uppercase\expandafter{\romannumeral7}, we provided a $Mathematica$ code to calculate the corresponding GBZ and auxiliary GBZ~\cite{SM1}.

\begin{figure}[t]
	\centerline{\includegraphics[height=4.35cm]{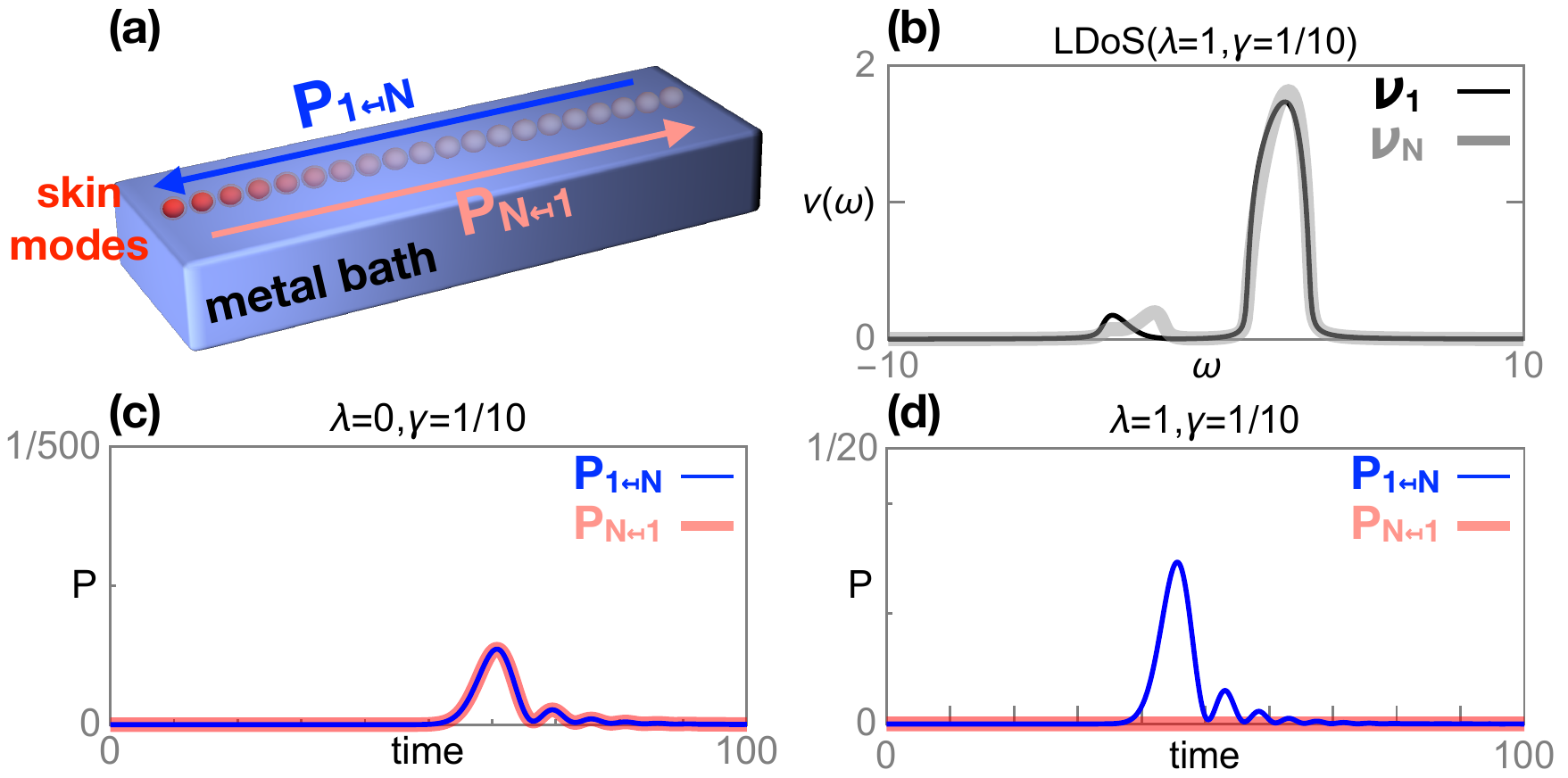}}
	\caption{LDoS and chiral tunneling effect induced by the skin modes. (a) shows the setup and the corresponding left localized skin modes in the model $\mathcal{H}_{\rm{spinless}}(k)-i\gamma\sigma_0$ with open boundary condition, where $t_1=2,t_2=\mu=1,\gamma=1/10,N=50$. (b) shows the LDoS at each boundary. There is no huge difference between them. (c) and (d) show the tunneling without and with skin modes, respectively. With the increasing of $\lambda$ in Eq.~4, skin modes emerge and the tunneling becomes chiral.
		\label{F3}}
\end{figure}

{\em Chiral tunneling effect.}---When looking for a proper physical observable for detecting skin modes, the local DoS (LDoS) may be the first physical quantity that comes into mind. However, we found that even if the skin modes are localized at one boundary, say, left boundary as shown in Fig.~\ref{F3} (a), it will not make the left LDoS much larger than the right one. As shown in Fig.~\ref{F3} (b), we plot the LDoS at each boundary of the  Bloch Hamiltonian $\mathcal{H}_{\rm{spinless}}(k)-i\gamma\sigma_0$ with open boundary condition (labeled by $H_{\rm{OBC}}$), where the black line ($\nu_1$) and gray line ($\nu_N$) represent the left and right LDoS, respectively. There is no huge difference between them. In the SM  \uppercase\expandafter{\romannumeral6}, we show that the LDoS at site $i$ can be expressed as
\begin{equation}
\nu_i(\omega)=-\frac{1}{\pi}\sum_n\Im \left[\frac{\langle i|\beta_{n}^R\rangle\langle \beta_{n}^L|i\rangle}{\omega-E_n}\right],
\label{E6}
\end{equation}	
where  $H_{OBC}|\beta_{n}^R\rangle=E_n|\beta_{n}^R\rangle$,  $H_{OBC}^\dag|\beta_{n}^L\rangle=E_n^*|\beta_{n}^L\rangle$, and $\langle \beta_{m}^L|\beta_{n}^R\rangle=\delta_{mn}$~\cite{Brody_2013}. We note that in the thermodynamic limit, $|\beta_{n}^R\rangle$ is a superposition of two non-Bloch waves with the same $|\beta_n|=r_n$~\cite{yokomizoNonBlochBandTheory2019a,2019arXiv191205499Y}. It can be further shown that $\langle i|\beta_{n}^R\rangle\propto r_n^i$ and $\langle \beta_{n}^L|i\rangle\propto 1/r_n^i$~\cite{SM1}. Therefore, the contribution of skin modes in Eq.~\ref{E6} cancels, which explains the numerical results of Fig.~\ref{F3} (b). Consequently, the LDoS is ineffective for detecting skin modes.

We now show that the existence of skin modes can be detected by the chiral tunneling effect due to the unidirectional nature of the non-Hermitian skin effects. This can be intuitively expected as the model with skin effect can be related to a model with non-reciprocal terms (which implies an asymmetric tunneling) by applying a proper basis (or gauge) transformation~\cite{2019arXiv191210048B}. As shown in Fig.~\ref{F3} (c)-(d), we plot $P_{N\leftarrow 1}(t)$ and $P_{1\leftarrow N}(t)$ of $H_{OBC}$ for different values of $\lambda$, where $P_{f\leftarrow i}(t)=|\langle f|U(t)|i\rangle|^2=|\langle f|e^{-i\hat{H}t}|i\rangle|^2$ is the tunneling strength from site $i$ to site $j$. When the TRS-breaking parameter $\lambda$ increase from zero to a nonzero value, skin modes emerge, in the meantime, $P_{1\leftarrow N}(t)$ increases and $P_{N\leftarrow 1}(t)$ decreases.  This means the tunneling along the direction in which the skin modes are localized is favored. Based on the non-Bloch theory, in the SM \uppercase\expandafter{\romannumeral6}, we show that $P_{N\leftarrow 1}(t)=|\langle N|U(t)|1\rangle|^2\propto r_n^{N-1}$ and $P_{1\leftarrow N}(t)=|\langle 1|U(t)|N\rangle|^2\propto r_n^{1-N}$, where $r_n$ represents the localization length of the skin mode $|\beta_n^R\rangle$.  This means the strength of the asymmetric tunneling exponentially depends on the localization length of skin modes~\cite{SM2}. We finally note that the chiral tunneling effect in our model can be experimental controlled by tuning the external magnetic field, which may be useful in the electronics studies.

{\em Discussions and conclusions.}---Our results can be applied to more general non-Hermitian Hamiltonians. For example, suppose that the on-site dissipation is a function of $k$, that is, $\mathcal{H}_{s,eff}(k)=\mathcal{H}_s(k)-i\gamma(k)\Gamma_0$, the results of Table.~\ref{T1} remain valid if $\gamma(k)$ is an even function. While if $\gamma(k)$ is an odd function, the results between commutative and anti-commutative in Table.~\ref{T1} will be interchanged. For the general case $\mathcal{H}_{s,eff}(k)=\mathcal{H}_s(k)+\Sigma(k)$, where $\Sigma(k)=\Sigma(\omega=0,k)$ is the self-energy correction at zero frequency~\cite{2017arXiv170805841K,PhysRevB.99.201107}, once the Hamiltonian has spinless TRS$^\dag$ ($\bar{\mathcal{T}}_+$) or IS ($\mathcal{P}$) or both, skin modes are absent except the case $G_{\bar{\mathcal{T}}_-,(\mathcal{P\bar{T}})_+}$~\cite{SM8}.

In summary, our results provide a new approach to realize and control skin modes by tunning the {\em Hermitian} Hamiltonian. On the theoretical side, our standard Green's function method paves the way for the study the novel physical responses induced by non-Hermitian skin modes. On the experimental side, we expect our models and the prediction of chiral tunneling effect can be realized and observed in various physical systems.

\begin{acknowledgments}
	Z. Yang thanks to the valuable discussion with Kai Zhang, Jiangping Hu, Andrei Bernevig, Chen Fang, Ching-Kai Chiu, Dong E Liu, Haiping Hu, Chunhui Liu, Shengshan Qin and  CongCong Le. The work is supported by the Ministry of Science and Technology of China 973 program (No. 2017YFA0303100), National Science Foundation of China (Grant No. NSFC-11888101, 1190020, 11534014, 11334012), and the Strategic Priority Research Program of CAS (Grant No.XDB07000000).
\end{acknowledgments}	

\bibliography{aGBZ}

\begin{thebibliography}{126}
\bibitem [{SM3()}]{SM3}%
  \BibitemOpen
  \href@noop {} {\bibinfo  {journal} {This means the Hermitian system
  $\mathcal{H}_s(k)$ is non-superconducting}\ }\BibitemShut {NoStop}%
\bibitem [{SM4()}]{SM4}%
  \BibitemOpen
\bibfield  {journal} {  }\href@noop {} {\bibinfo  {journal} {This is because
  $\Gamma_0$ is a diagonal matrix. As a result, we have
  $[\mathcal{P},\Gamma_0]=[\mathcal{P},\mathcal{T}]=0$, which breaks the three
  conditions of the spinful case in our main results}\ }\BibitemShut {NoStop}%
\bibitem [{SM5()}]{SM5}%
  \BibitemOpen
\bibfield  {journal} {  }\href@noop {} {\bibinfo  {journal} {We note that a
  system has $\mathcal{PT}$ symmetry does not imply the system has TRS or IS}\
  }\BibitemShut {NoStop}%
\bibitem [{GBZ()}]{GBZ_intro}%
  \BibitemOpen
\bibfield  {journal} {  }\href@noop {} {\bibinfo  {journal} {The GBZ condition,
  for example $|\beta_p(E)|=|\beta_{p+1}(E)|$, tells us that if for some
  $E_0\in\mathbb{C}$, $|\beta_p(E_0)|=|\beta_{p+1}(E_0)|$, then, $E_0$ is an
  asymptotic eigenvalue of the Hamiltonian with open boundary condition, where
  $\beta_i$ is the $i$th largest (ordered by the absolute values) root of
  $f(\beta,E)=0$ and $p$ is the order of the pole of $f(\beta,E)$}\
  }\BibitemShut {NoStop}%
\bibitem [{SM6()}]{SM6}%
  \BibitemOpen
\bibfield  {journal} {  }\href@noop {} {\bibinfo  {journal} {This is because
  for any given $E_0\in\mathbb{C}$, the symmetry constraints to the
  characteristic equation $f(\beta,E_0)=f(1/\beta,E_0)$ implies the roots must
  come in pairs as $(\beta,1/\beta)$. Combining this with GBZ condition
  $|\beta_p(E_0)|=|\beta_{p+1}(E_0)|$, we can have $|\beta_p|=|\beta_{p+1}|=1$,
  which indicates the absence of skin modes by recalling $\beta=e^{ik}$}\
  }\BibitemShut {NoStop}%
\bibitem [{SM7()}]{SM7}%
  \BibitemOpen
\bibfield  {journal} {  }\href@noop {} {\bibinfo  {journal} {This is because
  the GBZ condition is changed from $|\beta_p(E)|=|\beta_{p+1}(E)|$ to
  $|\beta_{p-1}(E)|=|\beta_p(E)|,$ $\&$ $|\beta_{p+1}(E)|=|\beta_{p+2}(E)|$. As
  a result, the symmetry constraint $f(\beta,E)=f(1/\beta,E)$ can no longer
  forbids the emergence of skin modes}\ }\BibitemShut {NoStop}%
\bibitem [{SM2()}]{SM2}%
  \BibitemOpen
  \href@noop {} {\bibinfo  {journal} {When $r_n<1$, it seems that the term
  $r_{n}^{1-N}$ will tend to infinity in the limit of $N\rightarrow\infty$.
  However, in the SM \uppercase\expandafter{\romannumeral6}, we show that the
  tunneling probability is bounded to less than 1, when the system is purely
  dissipative}\ }\BibitemShut {NoStop}%
\bibitem [{SM8()}]{SM8}%
  \BibitemOpen
\bibfield  {journal} {  }\href@noop {} {\bibinfo  {journal} {We note that this
  conclusion can only be applied to the cases in which $\mathcal{H}_{s,eff}(k)$
  has neither PHS nor PHS$^\dag$}\ }\BibitemShut {NoStop}%
\end{thebibliography}
\bibliographystyle{apsrev4-1}

\onecolumngrid
\newpage


\section*{Supplementary material}

# Supplemental Materials for
# "Non-Hermitian skin modes induced by on-site dissipations and chiral tunneling effect"

Yifei Yi[1,2] and Zhesen Yang[1,2]*

[1]*Beijing National Laboratory for Condensed Matter Physics,*
*and Institute of Physics, Chinese Academy of Sciences, Beijing 100190, China and*
[2]*University of Chinese Academy of Sciences, Beijing 100049, China*
(Dated: September 14, 2020)

## CONTENTS

---

* yangzs@iphy.ac.cn

## I. THE DERIVATION OF EQ. 2 IN THE MAIN TEXT

In this section, we will use the Green's function method to derive Eq. 2 from Eq. 1 in the main text. Start with the following one-dimensional (1D) *Hermitian* Hamiltonian,

$$
\begin{aligned}
\hat{H} &= \hat{H}_s + \hat{H}_b + \hat{H}_{s-b}, \\
\hat{H}_s &= \sum_{i,j} \sum_{\mu,\nu} t_{ij}^{\mu\nu} \hat{c}_{i\mu}^\dagger \hat{c}_{j\nu}, \\
\hat{H}_b &= \sum_{i,p_\mu,\mu} \left( \varepsilon_{p_\mu} - \mu_{p_\mu} \right) \hat{b}_{ip_\mu}^\dagger \hat{b}_{ip_\mu}, \\
\hat{H}_{s-b} &= \sum_{i,p_\mu,\mu} V_{p_\mu\mu} (\hat{c}_{i\mu}^\dagger \hat{b}_{ip_\mu} + \hat{b}_{ip_\mu}^\dagger \hat{c}_{i\mu}).
\end{aligned}
\tag{1}
$$

When the external bath degrees are integrated out, we can obtain the following Dyson equation of the retarded Green's function [1]:

$$
G_s^R(k,\omega) = [\omega - \mathcal{H}_s(k) - \Sigma_b^R(\omega)]^{-1},
$$

where

$$
[\mathcal{H}_s(k)]_{\mu\nu} = \sum_{l_{\mu\nu}} t_{l_{\mu\nu}}^{\mu\nu} e^{ikl_{\mu\nu}}
$$

is the Bloch Hamiltonian of the system and the diagonal matrix

$$
[\Sigma_b^R(\omega)]_{\mu\nu} = \delta_{\mu\nu} \sum_{p_\mu} |V_{p_\mu\mu}|^2 / (\omega - \varepsilon_{p_\mu} + \mu_{p_\mu} + i\eta)
$$

with $\eta = 0^+$ is the self-energy correction. The imaginary part of the self-energy correction corresponds to the spectral function of the external bath $[\Gamma(\omega)]_{\mu\nu} = \pi \delta_{\mu\nu} \sum_{p_\mu} |V_{p_\mu\mu}|^2 \delta\left(\omega - \varepsilon_{p_\mu} + \mu_{p_\mu}\right)$. A simple treatment of the dissipation is to assume an uniform distribution of $[\Gamma(\omega)]_{\mu\nu}$ in the region $[-W, W]$ [1]. If $2W$ is much larger than the band width of the system we concerned, $\Gamma(\omega)$ can be approximated by a constant diagonal matrix $i\gamma\Gamma_0$, and the corresponding non-Hermitian effective Bloch Hamiltonian can be written as

$$
\mathcal{H}_{s,eff}(k) = \mathcal{H}_s(k) - i\gamma\Gamma_0,
\tag{2}
$$

where $\gamma$ is proportional to the density of states (DoS) of the external bath and the system-bath coupling strength, and the diagonal matrix $\Gamma_0$ satisfies $\Gamma_0 = \Gamma_0^* = \Gamma_0^\dagger$. Thus, we obtained Eq.2 in the main text.

## II. SYMMETRY CONSTRAINT TO THE CHARACTERISTIC EQUATION

In this section, we will derive the symmetry constraint to the characteristic equation of the non-Bloch Hamiltonian $\mathcal{H}(\beta)$ [2–5]. Since all the elements of the non-Hermitian symmetry groups can be generated from the following five symmetries: time-reversal symmetry $\mathcal{T}$ (TRS), anomalous time-reversal symmetry $\tilde{\mathcal{T}}$ (TRS†), particle-hole symmetry $\mathcal{C}$ (PHS), anomalous particle-hole symmetry $\tilde{\mathcal{C}}$ (PHS†), and inversion symmetry $\mathcal{P}$ (IS), we here only provided the derivations of the above five symmetries.

## A. TRS

For the TRS, the constraint to the Bloch Hamiltonian with periodic boundary condition is

$$U_{\mathcal{T}}\mathcal{H}^*(-k)U_{\mathcal{T}}^{-1} = \mathcal{H}(k). \tag{3}$$

This implies the characteristic equation satisfying

$$f(k,E) = \det[E - \mathcal{H}(k)] = \det[E - U_{\mathcal{T}}\mathcal{H}^*(-k)U_{\mathcal{T}}^{-1}] = \det[E - \mathcal{H}^*(-k)]$$
$$= (\det[E^* - \mathcal{H}(-k)])^* = f^*(-k, E^*). \tag{4}$$

Now suppose that the characteristic equation can be expressed as

$$f(k,E) = \sum_{l,m,n} c_{l,m,n} E^l (\sin k)^m (\cos k)^n. \tag{5}$$

Then, according to Eq. 4

$$\sum_{l,m,n} c_{l,m,n} E^l (\sin k)^m (\cos k)^n = \sum_{l,m,n} c_{l,m,n}^* E^l (-\sin k)^m (\cos k)^n, \tag{6}$$

one can obtain,

$$c_{l,m,n} = (-1)^m c_{l,m,n}^*. \tag{7}$$

This means

$$\lambda_{l,2m_0,n} := c_{l,2m_0,n} \text{ is real}, \qquad i\lambda_{l,2m_0+1,n} := c_{l,2m_0+1,n} \text{ is imaginary}, \tag{8}$$

where $m_0$ is an integer. Now we extend $\beta = e^{ik}$ to the entire complex plane. The characteristic equation becomes

$$f(\beta, E) = \sum_{l,m_0,n} \left[ \lambda_{l,2m_0,n} E^l \left( \frac{\beta^2-1}{2i\beta} \right)^{2m_0} \left( \frac{\beta^2+1}{2\beta} \right)^n + i\lambda_{l,2m_0+1,n} E^l \left( \frac{\beta^2-1}{2i\beta} \right)^{2m_0+1} \left( \frac{\beta^2+1}{2\beta} \right)^n \right]$$
$$= \sum_{l,m_0,n} \left[ \lambda_{l,2m_0,n} (E^*)^l \left( \frac{(\beta^*)^2-1}{2i\beta^*} \right)^{2m_0} \left( \frac{(\beta^*)^2+1}{2\beta^*} \right)^n + i\lambda_{l,2m_0+1,n} (E^*)^l \left( \frac{(\beta^*)^2-1}{2i\beta^*} \right)^{2m_0+1} \left( \frac{(\beta^*)^2+1}{2\beta^*} \right)^n \right]^*$$
$$= [f(\beta^*, E^*)]^*. \tag{9}$$

Since $f(\beta, E) = 0 = [f(\beta, E)]^*$, we finally obtain

$$f(\beta, E) = f(\beta^*, E^*). \tag{10}$$

We note that our derivation requires the fact that the characteristic polynomial should be zero, i.e. $f(\beta, E) = \det[E - H(\beta)] = 0$. This condition ($f(\beta, E) = 0$) is allowed since the generalized Brillouin zone (GBZ) is totally determined by the characteristic equation $f(\beta, E) = 0$ and the corresponding GBZ condition [2–5], e.g., $|\beta_p(E)| = |\beta_{p+1}(E)|$, where $\beta_i$ is the $i$th largest (ordered by the absolute values) root of $f(\beta, E) = 0$, and $p$ is the order of the pole of $f(\beta, E) = 0$.

## B. TRS$^\dagger$

For the TRS$^\dagger$, the constraint to the Bloch Hamiltonian with periodic boundary condition is

$$U_{\mathcal{T}}\mathcal{H}^t(-k)U_{\mathcal{T}}^{-1} = \mathcal{H}(k). \tag{11}$$

This implies the characteristic equation satisfying

$$f(k,E) = \det[E - \mathcal{H}(k)] = \det[E - U_{\mathcal{T}}\mathcal{H}^t(-k)U_{\mathcal{T}}^{-1}] = \det[E - \mathcal{H}^t(-k)] = \det[E - \mathcal{H}(-k)] = f(-k, E). \tag{12}$$

Now suppose that the characteristic equation can be expressed as

$$f(k, E) = \sum_{l,m,n} c_{l,m,n} E^l (\sin k)^m (\cos k)^n. \tag{13}$$

The constraint of Eq. 12 implies

$$c_{l,m,n} = (-1)^m c_{l,m,n}, \tag{14}$$

which is equivalent to

$$c_{l,2m_0+1,n} = 0, \tag{15}$$

where $m_0$ is an integer. Now we extend $\beta = e^{ik}$ to the entire complex plane. The characteristic equation becomes

$$
\begin{aligned}
f(\beta, E) &= \sum_{l,m_0,n} c_{l,2m_0,n} E^l \left( \frac{\beta^2 - 1}{2i\beta} \right)^{2m_0} \left( \frac{\beta^2 + 1}{2\beta} \right)^n \\
&= \sum_{l,m_0,n} c_{l,2m_0,n} E^l \left( \frac{1 - \beta^2}{2i\beta} \right)^{2m_0} \left( \frac{\beta^2 + 1}{2\beta} \right)^n \\
&= f(1/\beta, E).
\end{aligned}
\tag{16}
$$

## C. PHS

For the PHS, the constraint to the Bloch Hamiltonian with periodic boundary condition is

$$-U_{\mathcal{C}} \mathcal{H}^t(-k) U_{\mathcal{C}}^{-1} = \mathcal{H}(k). \tag{17}$$

This implies the characteristic equation satisfying

$$f(k, E) = \det[E - \mathcal{H}(k)] = \det[E + U_{\mathcal{C}} \mathcal{H}^t(-k) U_{\mathcal{C}}^{-1}] = \det[E + \mathcal{H}(-k)] = (-1)^d \det[-E - \mathcal{H}(-k)] = (-1)^d f(-k, -E). \tag{18}$$

where $d$ is the dimension of the matrix. Now suppose that the characteristic equation can be expressed as

$$f(k, E) = \sum_{l,m,n} c_{l,m,n} E^l (\sin k)^m (\cos k)^n. \tag{19}$$

The constraint of Eq. 18 implies

$$c_{l,m,n} = (-1)^{d+l+m} c_{l,m,n}, \tag{20}$$

Now we extend $\beta = e^{ik}$ to the entire complex plane. The characteristic equation becomes

$$
\begin{aligned}
f(\beta, E) &= \sum_{l,m,n} c_{l,m,n} E^l \left( \frac{\beta^2 - 1}{2i\beta} \right)^m \left( \frac{\beta^2 + 1}{2\beta} \right)^n \\
&= \sum_{l,m,n} (-1)^{d+l+m} c_{l,m,n} E^l \left( \frac{\beta^2 - 1}{2i\beta} \right)^m \left( \frac{\beta^2 + 1}{2\beta} \right)^n \\
&= (-1)^d \sum_{l,m,n} c_{l,m,n} (-E)^l \left( \frac{1 - \beta^2}{2i\beta} \right)^m \left( \frac{\beta^2 + 1}{2\beta} \right)^n \\
&= (-1)^d f(1/\beta, -E).
\end{aligned}
\tag{21}
$$

However, in general, the matrix dimension of the Bloch Hamiltonian is even number, thus we finally obtain

$$f(\beta, E) = f(1/\beta, -E). \tag{22}$$

## D. PHS†

For the PHS†, the constraint to the Bloch Hamiltonian with periodic boundary condition is

$$-U_{\mathcal{C}}\mathcal{H}^*(-k)U_{\tilde{\mathcal{C}}}^{-1} = \mathcal{H}(k). \tag{23}$$

This implies the characteristic equation satisfying

$$
\begin{aligned}
f(k,E) &= \det[E - \mathcal{H}(k)] = \det[E + U_{\mathcal{C}}\mathcal{H}^*(-k)U_{\mathcal{C}^{-1}}] = \det[E + \mathcal{H}^*(-k)] \\
&= (-1)^d \det[-E - \mathcal{H}^*(-k)] = (-1)^d (\det[-E^* - \mathcal{H}(-k)])^* = (-1)^d f^*(-k, -E^*).
\end{aligned}
\tag{24}
$$

where $d$ is the dimension of the matrix. Now suppose that the characteristic equation can be expressed as

$$f(k,E) = \sum_{l,m,n} c_{l,m,n} E^l (\sin k)^m (\cos k)^n. \tag{25}$$

Then, according to Eq. 24

$$\sum_{l,m,n} c_{l,m,n} E^l (\sin k)^m (\cos k)^n = (-1)^d \sum_{l,m,n} c_{l,m,n}^* (-E)^l (-\sin k)^m (\cos k)^n, \tag{26}$$

one can obtain,

$$c_{l,m,n} = (-1)^{d+l+m} c_{l,m,n}^*. \tag{27}$$

Now we extend $\beta = e^{ik}$ to the entire complex plane. The characteristic equation becomes

$$
\begin{aligned}
f(\beta, E) &= \sum_{l,m,n} c_{l,m,n} E^l \left( \frac{\beta^2 - 1}{2i\beta} \right)^m \left( \frac{\beta^2 + 1}{2\beta} \right)^n \\
&= \sum_{l,m,n} (-1)^{d+l+m} c_{l,m,n}^* E^l \left( \frac{\beta^2 - 1}{2i\beta} \right)^m \left( \frac{\beta^2 + 1}{2\beta} \right)^n \\
&= (-1)^d \sum_{l,m,n} c_{l,m,n}^* (-E)^l \left( -\frac{\beta^2 - 1}{2i\beta} \right)^m \left( \frac{\beta^2 + 1}{2\beta} \right)^n \\
&= (-1)^d \sum_{l,m,n} \left[ c_{l,m,n} (-E^*)^l \left( \frac{(\beta^*)^2 - 1}{2i\beta^*} \right)^m \left( \frac{(\beta^*)^2 + 1}{2\beta^*} \right)^n \right]^* \\
&= (-1)^d [f(\beta^*, -E^*)]^*.
\end{aligned}
\tag{28}
$$

Since $f(\beta, E) = (-1)^d [f(\beta^*, -E^*)]^* = 0$, we finally obtain

$$f(\beta, E) = f(\beta^*, -E^*). \tag{29}$$

## E. IS

For the IS, the constraint to the Bloch Hamiltonian with periodic boundary condition is

$$U_{\mathcal{P}}\mathcal{H}(-k)U_{\mathcal{P}}^{-1} = \mathcal{H}(k). \tag{30}$$

This implies the characteristic equation satisfying

$$f(k,E) = \det[E - \mathcal{H}(k)] = \det[E - U_{\mathcal{P}}\mathcal{H}(-k)U_{\mathcal{P}}^{-1}] = \det[E - \mathcal{H}(-k)] = f(-k, E). \tag{31}$$

Now suppose that the characteristic equation can be expressed as

$$f(k,E) = \sum_{l,m,n} c_{l,m,n} E^l (\sin k)^m (\cos k)^n. \tag{32}$$

The constraint of Eq. 31 implies

$$c_{l,m,n} = (-1)^m c_{l,m,n}, \tag{33}$$

which is equivalent to

$$c_{l,2m_0+1,n} = 0, \tag{34}$$

where $m$ is an integer. Now we extend $\beta = e^{ik}$ to the entire complex plane. The characteristic equation becomes

$$
\begin{aligned}
f(\beta, E) &= \sum_{l,m_0,n} c_{l,2m_0,n} E^l \left( \frac{\beta^2 - 1}{2i\beta} \right)^{2m_0} \left( \frac{\beta^2 + 1}{2\beta} \right)^n \\
&= \sum_{l,m_0,n} c_{l,2m_0,n} E^l \left( \frac{1 - \beta^2}{2i\beta} \right)^{2m_0} \left( \frac{\beta^2 + 1}{2\beta} \right)^n \\
&= f(1/\beta, E).
\end{aligned}
\tag{35}
$$

## III. DERIVATION OF THE NON-HERMITIAN SYMMETRY GROUPS WE CONCERNED

In the main text, we have mentioned that if $\mathcal{H}_s(k)$ preserves TRS but breaks PHS, $\mathcal{H}_{s,eff}(k)$ belongs to the following non-Hermitian symmetry groups $G_{\mathcal{T}_\pm}$, $G_{\mathcal{T}_\pm,(\mathcal{PT})_\pm}$, $G_{\mathcal{T}_\pm,(\mathcal{PC})_\pm}$, $G_{\mathcal{T}_\pm,(\mathcal{P\bar{T}})_\pm}$, $G_{\mathcal{T}_\pm,(\mathcal{P\bar{C}})_\pm}$, $G_{\bar{\mathcal{T}}_\pm}$, $G_{\bar{\mathcal{T}}_\pm,(\mathcal{PT})_\pm}$, $G_{\bar{\mathcal{T}}_\pm,(\mathcal{PC})_\pm}$, $G_{\bar{\mathcal{T}}_\pm,(\mathcal{P\bar{T}})_\pm}$, $G_{\bar{\mathcal{T}}_\pm,(\mathcal{P\bar{C}})_\pm}$. In this section, we will show this.

### A. Hermitian case

TABLE I. Hermitian symmetry classes. There exist three independent symmetry generators, TRS ($\mathcal{T}$), PHS ($\mathcal{C}$), and IS ($\mathcal{P}$). Here the red colors represent the Hermitian symmetry groups we concerned, which preserve TRS but break PHS.

| Group generators | Symmetries | Group generators | Symmetries |
|---|---|---|---|
| I | $U_I = \mathcal{I}$ | $\mathcal{P} = U_{\mathcal{P}}$ | $U_{\mathcal{P}}^2 = 1$ |
| $\mathcal{T} = U_{\mathcal{T}} \mathcal{K}^*$ | $U_{\mathcal{T}} U_{\mathcal{T}}^* = \pm 1$ | $\mathcal{PT} = U_{\mathcal{PT}} \mathcal{K}^*$ | $U_{\mathcal{PT}} U_{\mathcal{PT}}^* = \pm 1$ |
| $\mathcal{C} = U_{\mathcal{C}} \mathcal{K}^t$ | $U_{\mathcal{C}} U_{\mathcal{C}}^* = \pm 1$ | $\mathcal{PC} = U_{\mathcal{PC}} \mathcal{K}^t$ | $U_{\mathcal{PC}} U_{\mathcal{PC}}^* = \pm 1$ |
| $\mathcal{TC} = U_{\mathcal{TC}} \mathcal{K}^\dagger$ | $U_{\mathcal{TC}}^2 = 1$ | $\mathcal{PTC} = U_{\mathcal{PTC}} \mathcal{K}^\dagger$ | $(U_{\mathcal{PTC}})^2 = 1$ |

| Group generators | Symmetries |
|---|---|
| $\{\mathcal{P} = U_{\mathcal{P}}, \mathcal{TC} = U_{\mathcal{TC}} \mathcal{K}^\dagger\}$ | $(U_{\mathcal{P}} U_{\mathcal{TC}})^2 = \pm 1$ |
| $\{\mathcal{P} = U_{\mathcal{P}}, \mathcal{T} = U_{\mathcal{T}} \mathcal{K}^*\}$ | $U_{\mathcal{T}} U_{\mathcal{T}}^* = \pm 1, (U_{\mathcal{P}} U_{\mathcal{T}})(U_{\mathcal{P}} U_{\mathcal{T}})^* = \pm 1$ |
| $\{\mathcal{P} = U_{\mathcal{P}}, \mathcal{C} = U_{\mathcal{C}} \mathcal{K}^t\}$ | $U_{\mathcal{C}} U_{\mathcal{C}}^* = \pm 1, (U_{\mathcal{P}} U_{\mathcal{C}})(U_{\mathcal{P}} U_{\mathcal{C}})^* = \pm 1$ |
| $\{\mathcal{T} = U_{\mathcal{T}} \mathcal{K}^*, \mathcal{C} = U_{\mathcal{C}} \mathcal{K}^t\}$ | $U_{\mathcal{T}} U_{\mathcal{T}}^* = \pm 1, U_{\mathcal{C}} U_{\mathcal{C}}^* = \pm 1$ |
| $\{\mathcal{T} = U_{\mathcal{T}} \mathcal{K}^*, \mathcal{PC} = U_{\mathcal{PC}} \mathcal{K}^t\}$ | $U_{\mathcal{T}} U_{\mathcal{T}}^* = \pm 1, U_{\mathcal{PC}} U_{\mathcal{PC}}^* = \pm 1$ |
| $\{\mathcal{PT} = U_{\mathcal{PT}} \mathcal{K}^*, \mathcal{C} = U_{\mathcal{C}} \mathcal{K}^t\}$ | $U_{\mathcal{PT}} U_{\mathcal{PT}}^* = \pm 1, U_{\mathcal{C}} U_{\mathcal{C}}^* = \pm 1$ |
| $\{\mathcal{PT} = U_{\mathcal{PT}} \mathcal{K}^*, \mathcal{PC} = U_{\mathcal{PC}} \mathcal{K}^t\}$ | $U_{\mathcal{PT}} U_{\mathcal{PT}}^* = \pm 1, U_{\mathcal{PC}} U_{\mathcal{PC}}^* = \pm 1$ |

| Group generators | Symmetries |
|---|---|
| $\{\mathcal{P} = U_{\mathcal{P}}, \mathcal{T} = U_{\mathcal{T}} \mathcal{K}^*, \mathcal{C} = U_{\mathcal{C}} \mathcal{K}^t\}$ | $U_{\mathcal{T}} U_{\mathcal{T}}^* = \pm 1, U_{\mathcal{C}} U_{\mathcal{C}}^* = \pm 1, (U_{\mathcal{P}} U_{\mathcal{T}})(U_{\mathcal{P}} U_{\mathcal{T}})^* = \pm 1, (U_{\mathcal{P}} U_{\mathcal{C}})(U_{\mathcal{P}} U_{\mathcal{C}})^* = \pm 1$ |

In 1D Hermitian system, there exist three non-spatial symmetries, TRS, PHS, and chiral symmetry (CS), and one spatial symmetry, IS. Here CS is a combination of TRS and PHS. Thus, the independent symmetry generators are TRS, PHS and IS. Based on these three independent generators, other symmetries can be obtained by combining two or three of them, that is

$$\mathcal{T}, \mathcal{C}, \mathcal{P} \quad \rightarrow \quad \mathcal{TC}, \mathcal{PT}, \mathcal{PC}, \mathcal{PTC}. \tag{36}$$

Note that in the Hermitian case, since the transpose operator $\mathcal{K}^t$ and complex conjugate operator $\mathcal{K}^*$ are equivalent due to $\mathcal{H} = \mathcal{H}^\dagger$, the PHS, which is defined by the transpose operator $\mathcal{C} = U_\mathcal{C}\mathcal{K}^t$, is equivalent to $\mathcal{C} = U_\mathcal{C}\mathcal{K}^*$ [6, 7]. Mathematically, the role of Hermitian symmetries can be classified by

$$\text{(i)}: \ k \to (k, -k), \qquad \text{(ii)}: \ E \to (E, -E), \qquad \text{(iii)}: \ \mathcal{H} \to (\mathcal{H} = \mathcal{H}^\dagger, \mathcal{H}^* = \mathcal{H}^t), \tag{37}$$

which implies there only exist $2 \times 2 \times 2 = 8$ group elements in all 1D Hermitian groups, including the 7 symmetries listed in Eq. 36 and the identity element I.

Now, for the Hermitian case, all the symmetries can be generated by the following independent generators

$$\begin{aligned}
\text{TRS} \ (\mathcal{T} = U_\mathcal{T}\mathcal{K}^*): \quad & U_\mathcal{T}\mathcal{H}^*(k)U_\mathcal{T}^{-1} = \mathcal{H}(-k), \qquad U_\mathcal{T}U_\mathcal{T}^* = \pm 1, \\
\text{PHS} \ (\mathcal{C} = U_\mathcal{C}\mathcal{K}^t): \quad & U_\mathcal{C}\mathcal{H}^t(k)U_\mathcal{C}^{-1} = -\mathcal{H}(-k), \quad U_\mathcal{C}U_\mathcal{C}^* = \pm 1, \\
\text{IS} \quad (\mathcal{P} = U_\mathcal{P}): \quad & U_\mathcal{P}\mathcal{H}(k)U_\mathcal{P}^{-1} = \mathcal{H}(-k), \qquad U_\mathcal{P}^2 = 1.
\end{aligned} \tag{38}$$

The complete symmetry classification of 1D Hermitian Hamiltonian are shown in Table. I. Here we note that

$$U_{\mathcal{P}X}U_{\mathcal{P}X}^* = -pU_XU_X^*, \quad [\mathcal{P}, X]_p = 0, \tag{39}$$

where $X = \mathcal{T}, \mathcal{C}$ and $p = \pm$. Here is the proof. Without loss of generality, we let $X = \mathcal{T}$.

- If $p = -1$, namely, $[\mathcal{P}, \mathcal{T}] = [\mathcal{P}, \mathcal{T}]_- = 0$, we have $\mathcal{PT} - \mathcal{TP} = 0 \ \to \ \mathcal{T} = \mathcal{PTP} \ \to \ e^{i\theta}U_\mathcal{T}\mathcal{K}^* = U_\mathcal{P}e^{i\theta}U_\mathcal{T}\mathcal{K}^*U_\mathcal{P} \ \to \ U_\mathcal{T} = U_\mathcal{P}U_\mathcal{T}U_\mathcal{P}^* \ \to \ U_\mathcal{T}U_\mathcal{T}^* = U_\mathcal{P}U_\mathcal{T}U_\mathcal{P}^*U_\mathcal{T}^* = (U_\mathcal{P}U_\mathcal{T})(U_\mathcal{P}U_\mathcal{T})^* = U_{\mathcal{PT}}U_{\mathcal{PT}}^*$.

- If $p = +1$, namely, $\{\mathcal{P}, \mathcal{T}\} = [\mathcal{P}, \mathcal{T}]_+ = 0$, we have $\mathcal{PT} + \mathcal{TP} = 0 \ \to \ \mathcal{T} = -\mathcal{PTP} \ \to \ e^{i\theta}U_\mathcal{T}\mathcal{K}^* = -U_\mathcal{P}e^{i\theta}U_\mathcal{T}\mathcal{K}^*U_\mathcal{P} \ \to \ U_\mathcal{T} = -U_\mathcal{P}U_\mathcal{T}U_\mathcal{P}^* \ \to \ U_\mathcal{T}U_\mathcal{T}^* = -U_\mathcal{P}U_\mathcal{T}U_\mathcal{P}^*U_\mathcal{T}^* = -(U_\mathcal{P}U_\mathcal{T})(U_\mathcal{P}U_\mathcal{T})^* = -U_{\mathcal{PT}}U_{\mathcal{PT}}^*$.

If $X = \mathcal{C}$, the proof are the same by using the fact $U_\mathcal{C}^* = U_\mathcal{C}^t$.

According to Table. I, there exists 54 symmetry classes for the 1D Hermitian Hamiltonians. As discussed in the main text, we only focus on the systems preserving TRS but breaking PHS. According to Table. I, only the symmetry classes with red color will be discussed in the following section, namely,

$$G_{\mathcal{T}_\pm}, \quad G_{\mathcal{T}_\pm, (\mathcal{PT})_\pm}, \quad G_{\mathcal{T}_\pm, (\mathcal{PC})_\pm}. \tag{40}$$

We note that $\mathcal{P}$ and $\mathcal{PTC}$ are also group elements of $G_{\mathcal{T}_\pm, (\mathcal{PT})_\pm}$ and $G_{\mathcal{T}_\pm, (\mathcal{PC})_\pm}$, respectively. For example, the group $G_{\mathcal{T}_-, (\mathcal{PT})_+}$ contains the following symmetries, $\mathcal{T}_-$, $\mathcal{P}$, and $(\mathcal{PT})_+$. And according to Eq. 39, the following commutation relation $\{\mathcal{P}, \mathcal{T}\} = 0$ holds for $G_{\mathcal{T}_-, (\mathcal{PT})_+}$.

## B. Symmetry ramifications

Once the on-site dissipations are added to the Hermitian Hamiltonian $\mathcal{H}_s(k)$ in Eq. 2, the symmetries will be ramified. In this subsection, we will derive the ramification rule.

### 1. TRS

When the Hermitian system has TRS, which is represented by $\mathcal{T} = U_\mathcal{T}\mathcal{K}^*$, it must also preserve TRS$^\dagger$, which is represented by $\bar{\mathcal{T}} = U_\mathcal{T}\mathcal{K}^t$. The constraints to the Bloch Hamiltonian are

$$U_\mathcal{T}\mathcal{H}_s^*(k)U_\mathcal{T}^{-1} = U_\mathcal{T}\mathcal{H}_s^t(k)U_\mathcal{T}^{-1} = \mathcal{H}_s(-k). \tag{41}$$

Now, if we add the on-site dissipation to the Hermitian Hamiltonian,

$$\mathcal{H}_{s,eff}(k) = \mathcal{H}_s(k) + i\gamma\Gamma_0, \tag{42}$$

then, according to $\Gamma_0 = \Gamma_0^* = \Gamma_0^t$, the discussion can be classified by the commutation relation.

- $[U_\mathcal{T}, \Gamma_0] = 0$:

  On the one hand, $U_\mathcal{T}\mathcal{H}_{s,eff}^*(k)U_\mathcal{T}^{-1} = U_\mathcal{T}[\mathcal{H}_s(k) + i\gamma\Gamma_0]^*U_\mathcal{T}^{-1} = U_\mathcal{T}[\mathcal{H}_s^*(k) - i\gamma\Gamma_0^*]U_\mathcal{T}^{-1} = \mathcal{H}_s(-k) - i\gamma\Gamma_0 \neq \mathcal{H}_{s,eff}(-k)$. This means TRS is broken by adding on-site dissipation.

  On the other hand, $U_\mathcal{T}\mathcal{H}_{s,eff}^t(k)U_\mathcal{T}^{-1} = U_\mathcal{T}[\mathcal{H}_s(k) + i\gamma\Gamma_0]^tU_\mathcal{T}^{-1} = U_\mathcal{T}[\mathcal{H}_s^t(k) + i\gamma\Gamma_0^t]U_\mathcal{T}^{-1} = \mathcal{H}_s(-k) + i\gamma\Gamma_0 = \mathcal{H}_{s,eff}(-k)$. This means TRS$^\dagger$ is preserved by adding on-site dissipation.

- $\{U_\mathcal{T}, \Gamma_0\} = 0$:

  On the one hand, $U_\mathcal{T}\mathcal{H}_{s,eff}^*(k)U_\mathcal{T}^{-1} = U_\mathcal{T}[\mathcal{H}_s(k) + i\gamma\Gamma_0]^*U_\mathcal{T}^{-1} = U_\mathcal{T}[\mathcal{H}_s^*(k) - i\gamma\Gamma_0^*]U_\mathcal{T}^{-1} = \mathcal{H}_s(-k) + i\gamma\Gamma_0 = \mathcal{H}_{s,eff}(-k)$. This means TRS is preserved by adding on-site dissipation.

  On the other hand, $U_\mathcal{T}\mathcal{H}_{s,eff}^t(k)U_\mathcal{T}^{-1} = U_\mathcal{T}[\mathcal{H}_s(k) + i\gamma\Gamma_0]^tU_\mathcal{T}^{-1} = U_\mathcal{T}[\mathcal{H}_s^t(k) + i\gamma\Gamma_0^t]U_\mathcal{T}^{-1} = \mathcal{H}_s(-k) - i\gamma\Gamma_0 \neq \mathcal{H}_{s,eff}(-k)$. This means TRS$^\dagger$ is broken by adding on-site dissipation.

### 2. PHS

When the Hermitian system has PHS, which is represented by $\mathcal{C} = U_\mathcal{C}\mathcal{K}^t$, it must also preserve PHS$^\dagger$, which is represented by $\bar{\mathcal{C}} = U_\mathcal{C}\mathcal{K}^*$. The constraints to the Bloch Hamiltonian are

$$U_\mathcal{C}\mathcal{H}_s^*(k)U_\mathcal{C}^{-1} = U_\mathcal{C}\mathcal{H}_s^t(k)U_\mathcal{C}^{-1} = -\mathcal{H}_s(-k). \tag{43}$$

Now, if we add the on-site dissipation to the Hermitian Hamiltonian,

$$\mathcal{H}_{s,eff}(k) = \mathcal{H}_s(k) + i\gamma\Gamma_0, \tag{44}$$

then, according to $\Gamma_0 = \Gamma_0^* = \Gamma_0^t$, the discussion can be classified by the commutation relation.

- $[U_\mathcal{C}, \Gamma_0] = 0$:

  On the one hand, $U_\mathcal{C}\mathcal{H}_{s,eff}^t(k)U_\mathcal{C}^{-1} = U_\mathcal{C}[\mathcal{H}_s(k) + i\gamma\Gamma_0]^tU_\mathcal{C}^{-1} = U_\mathcal{C}[\mathcal{H}_s^t(k) + i\gamma\Gamma_0^t]U_\mathcal{C}^{-1} = -\mathcal{H}_s(-k) + i\gamma\Gamma_0 \neq -\mathcal{H}_{s,eff}(-k)$. This means PHS is broken by adding on-site dissipation.

  On the other hand, $U_\mathcal{C}\mathcal{H}_{s,eff}^*(k)U_\mathcal{C}^{-1} = U_\mathcal{C}[\mathcal{H}_s(k) + i\gamma\Gamma_0]^*U_\mathcal{C}^{-1} = U_\mathcal{C}[\mathcal{H}_s^*(k) - i\gamma\Gamma_0^*]U_\mathcal{C}^{-1} = -\mathcal{H}_s(-k) - i\gamma\Gamma_0 = -\mathcal{H}_{s,eff}(-k)$. This means PHS$^\dagger$ is preserved by adding on-site dissipation.

- $\{U_\mathcal{C}, \Gamma_0\} = 0$:

  On the one hand, $U_\mathcal{C}\mathcal{H}_{s,eff}^t(k)U_\mathcal{C}^{-1} = U_\mathcal{C}[\mathcal{H}_s(k) + i\gamma\Gamma_0]^tU_\mathcal{C}^{-1} = U_\mathcal{C}[\mathcal{H}_s^t(k) + i\gamma\Gamma_0^t]U_\mathcal{C}^{-1} = -\mathcal{H}_s(-k) - i\gamma\Gamma_0 = -\mathcal{H}_{s,eff}(-k)$. This means PHS is preserved by adding on-site dissipation.

  On the other hand, $U_\mathcal{C}\mathcal{H}_{s,eff}^*(k)U_\mathcal{C}^{-1} = U_\mathcal{C}[\mathcal{H}_s(k) + i\gamma\Gamma_0]^*U_\mathcal{C}^{-1} = U_\mathcal{C}[\mathcal{H}_s^*(k) - i\gamma\Gamma_0^*]U_\mathcal{C}^{-1} = -\mathcal{H}_s(-k) + i\gamma\Gamma_0 \neq -\mathcal{H}_{s,eff}(-k)$. This means PHS$^\dagger$ is broken by adding on-site dissipation.

### 3. Other symmetries

For other symmetries, if they map $E \to E$, which include I, $\mathcal{P}$, $\mathcal{PT}$, $\mathcal{T}$, they are equivalent to the case of $\mathcal{T}$. To be more precise, when $[\Gamma_0, U_X] = 0$, the following symmetries without $\mathcal{K}^*$ and $\mathcal{K}^\dagger$ are preserved, namely, I, $\mathcal{P}\bar{\mathcal{T}}$, $\mathcal{P}$, $\bar{\mathcal{T}}$. When $\{\Gamma_0, U_X\} = 0$, the following symmetries with $\mathcal{K}^*$ and $\mathcal{K}^\dagger$ are preserved, namely, $\mathcal{T}\bar{\mathcal{T}}$, $\mathcal{PT}$, $\mathcal{PT}\bar{\mathcal{T}}$, $\mathcal{T}$. On the other hand, if the symmetries map $E \to -E$, which include $\mathcal{TC}$, $\mathcal{PC}$, $\mathcal{PTC}$, $\mathcal{C}$, they are equivalent to the case of $\mathcal{C}$. To be more precise, when $[\Gamma_0, U_X] = 0$, the following symmetries with $\mathcal{K}^*$ and $\mathcal{K}^\dagger$ are preserved, namely, $\mathcal{TC}$, $\mathcal{P}\bar{\mathcal{C}}$, $\bar{\mathcal{C}}$, $\mathcal{PTC}$. When $\{\Gamma_0, U_X\} = 0$, the following symmetries without $\mathcal{K}^*$ and $\mathcal{K}^\dagger$ are preserved, namely, $\bar{\mathcal{T}}\mathcal{C}$, $\mathcal{PC}$, $\mathcal{C}$, $\mathcal{P}\bar{\mathcal{T}}\mathcal{C}$. These are the results listed in Table. 1.

### C. Non-Hermitian case

Based on the ramification rule, all the non-Hermitian symmetry groups can be derived. Here we note that the square relation can not be changed by adding the non-Hermitian term. For example, consider a Hermitian system with TRS, e.g. $\mathcal{T} = U_\mathcal{T}\mathcal{K}^*$ and $U_\mathcal{T}U_\mathcal{T}^* = -1$, then, if $[\Gamma_0, U_\mathcal{T}] = 0$, the non-Hermitian system must have TRS$^\dagger$ with $\bar{\mathcal{T}} = U_\mathcal{T}\mathcal{K}^t$ and $U_\mathcal{T}U_\mathcal{T}^* = -1$; if $\{\Gamma_0, U_\mathcal{T}\} = 0$, the TRS is preserved in non-Hermitian systems with $\mathcal{T} = U_\mathcal{T}\mathcal{K}^*$ and $U_\mathcal{T}U_\mathcal{T}^* = -1$. Therefore, with the existence of on-site dissipation, besides $G_{\mathcal{T}_\pm}$, $G_{\mathcal{T}_\pm, (\mathcal{PT})_\pm}$, $G_{\mathcal{T}_\pm, (\mathcal{PC})_\pm}$, the following 26 non-Hermitian symmetry groups can be ramified

$$G_{\bar{\mathcal{T}}_\pm}, G_{\mathcal{T}_\pm, (\mathcal{P}\bar{\mathcal{T}})_\pm}, G_{\bar{\mathcal{T}}_\pm, (\mathcal{PT})_\pm}, G_{\bar{\mathcal{T}}_\pm, (\mathcal{P}\bar{\mathcal{T}})_\pm}, G_{\mathcal{T}_\pm, (\mathcal{P}\bar{\mathcal{C}})_\pm}, G_{\bar{\mathcal{T}}_\pm, (\mathcal{PC})_\pm}, G_{\bar{\mathcal{T}}_\pm, (\mathcal{P}\bar{\mathcal{C}})_\pm}.$$

We finally note that the following non-Hermitian symmetries connected by "=" are equivalent.

$$\begin{aligned}
&\text{I} = \mathcal{T}\bar{\mathcal{T}}\mathcal{C}\bar{\mathcal{C}}, \quad \mathcal{T} = \bar{\mathcal{T}}\mathcal{C}\bar{\mathcal{C}}, \quad \bar{\mathcal{T}} = \mathcal{T}\mathcal{C}\bar{\mathcal{C}}, \quad \mathcal{C} = \mathcal{T}\bar{\mathcal{T}}\bar{\mathcal{C}}, \quad \bar{\mathcal{C}} = \mathcal{T}\bar{\mathcal{T}}\mathcal{C}, \quad \mathcal{P} = \mathcal{P}\mathcal{T}\bar{\mathcal{T}}\mathcal{C}\bar{\mathcal{C}} \\
&\mathcal{T}\bar{\mathcal{T}} = \mathcal{C}\bar{\mathcal{C}}, \quad \mathcal{T}\mathcal{C} = \bar{\mathcal{T}}\bar{\mathcal{C}}, \quad \bar{\mathcal{T}}\mathcal{C} = \mathcal{T}\bar{\mathcal{C}}, \quad \mathcal{PT} = \mathcal{P}\bar{\mathcal{T}}\mathcal{C}\bar{\mathcal{C}}, \quad \mathcal{P}\bar{\mathcal{T}} = \mathcal{P}\mathcal{T}\mathcal{C}\bar{\mathcal{C}}, \quad \mathcal{PC} = \mathcal{P}\mathcal{T}\bar{\mathcal{T}}\bar{\mathcal{C}}, \quad \mathcal{P}\bar{\mathcal{C}} = \mathcal{P}\mathcal{T}\bar{\mathcal{T}}\mathcal{C}, \\
&\mathcal{P}\mathcal{T}\bar{\mathcal{T}} = \mathcal{P}\mathcal{C}\bar{\mathcal{C}}, \quad \mathcal{PTC} = \mathcal{P}\bar{\mathcal{T}}\bar{\mathcal{C}}, \quad \mathcal{PT}\bar{\mathcal{C}} = \mathcal{P}\bar{\mathcal{T}}\mathcal{C}.
\end{aligned} \tag{45}$$

### D. An application example of Table. I in the main text

To make Table. I easy to understand, we employ the SSH model as a concrete example to illustrate the usage of Table. I. The Bloch Hamiltonian of the SSH model can be written as:

$$\mathcal{H}_{\mathrm{SSH}}(k) = (t_1 + t_2 \cos k)\sigma_x + t_2 \sin k \sigma_y \tag{46}$$

which preserves

$$\begin{aligned}
&\mathcal{T}_+ = \mathcal{K}^*, \mathcal{C}_+ = \sigma_z \mathcal{K}^t, \mathcal{TC} = \sigma_z \mathcal{K}^\dagger, \\
&\mathcal{P} = \sigma_x, (\mathcal{PT})_+ = \sigma_x \mathcal{K}^*, (\mathcal{PC})_- = \sigma_y \mathcal{K}^t, \mathcal{PTC} = \sigma_y \mathcal{K}^\dagger.
\end{aligned} \tag{47}$$

One can verify this result without any difficulty. For example, $\mathcal{T}_+ \mathcal{H}_{\mathrm{SSH}}(k)\mathcal{T}_+^{-1} = \mathcal{H}_{\mathrm{SSH}}^*(k) = \mathcal{H}_{\mathrm{SSH}}(-k)$ and $\mathcal{C}_+ \mathcal{H}_{\mathrm{SSH}}(k)\mathcal{C}_+^{-1} = \sigma_z \mathcal{H}_{\mathrm{SSH}}^t(k)\sigma_z = -\mathcal{H}_{\mathrm{SSH}}(-k)$.

Now, we add an on-site dissipation to the system. Consequently, the Bloch Hamiltonian becomes:

$$\mathcal{H}_{\mathrm{nH-SSH}}(k) = (t_1 + t_2 \cos k)\sigma_x + t_2 \sin k \sigma_y + i\gamma\sigma_z. \tag{48}$$

Since

$$\begin{aligned}
&[U_{\mathcal{T}_+}, \sigma_z] = [U_{\mathcal{C}_+}, \sigma_z] = [U_{\mathcal{TC}}, \sigma_z] = 0, \\
&\{U_\mathcal{P}, \sigma_z\} = \{U_{(\mathcal{PT})_+}, \sigma_z\} = \{U_{(\mathcal{PC})_-}, \sigma_z\} = \{U_{\mathcal{PTC}}, \sigma_z\} = 0,
\end{aligned} \tag{49}$$

according to the second and fifth rows of Table. I, the above non-Hermitian SSH model preserves the following non-Hermitian symmetries:

$$\begin{aligned}
&\bar{\mathcal{T}}_+ = \mathcal{K}^t, \bar{\mathcal{C}}_+ = \sigma_z \mathcal{K}^*, \mathcal{TC} = \sigma_z \mathcal{K}^\dagger, \\
&\mathcal{PT\bar{T}} = \sigma_x \mathcal{K}^\dagger, \mathcal{PT}_+ = \sigma_x \mathcal{K}^*, \mathcal{PC}_- = \sigma_y \mathcal{K}^t, \mathcal{PTC} = \sigma_y.
\end{aligned} \tag{50}$$

For example, $\bar{\mathcal{T}}_+ \mathcal{H}(k)\bar{\mathcal{T}}_+^{-1} = \mathcal{H}^t(k) = \mathcal{H}(-k)$, and $\bar{\mathcal{C}}_+ \mathcal{H}(k)\bar{\mathcal{C}}_+^{-1} = \sigma_z \mathcal{H}^*(k)\sigma_z = -\mathcal{H}(-k)$. Again, according to the fourth and seventh rows of Table. I, one can check that the characteristic equation of the non-Bloch Hamiltonian $f(\beta, E)$ satisfies:

$$\begin{aligned}
&f(\beta, E) \overset{\bar{\mathcal{T}}_+}{=} f(1/\beta, E), \quad f(\beta, E) \overset{\bar{\mathcal{C}}_+}{=} f(\beta^*, -E^*), \quad f(\beta, E) \overset{\mathcal{TC}}{=} f(1/\beta^*, -E^*), \quad f(\beta, E) \overset{\mathcal{PT\bar{T}}}{=} f(\beta^*, E^*), \\
&f(\beta, E) \overset{\mathcal{PT}_+}{=} f(1/\beta^*, E^*), \quad f(\beta, E) \overset{\mathcal{PC}_-}{=} f(\beta, -E), \quad f(\beta, E) \overset{\mathcal{PTC}}{=} f(1/\beta, -E).
\end{aligned} \tag{51}$$

Thus, we've shown how the Table. I is applied. In short, we start with a Hermitian Hamiltonian preserves certain Hermitian symmetries, then, the on-site dissipation terms are added. When the added terms are commute or anticommute with the presentations of these Hermitian symmetries, one can refer to the Table. I in the main text, and learn the non-Hermitian symmetries the corresponding non-Hermitian Hamiltonian preserves and the restrictions imposed on the characteristic equation of the non-Bloch Hamiltonian.

## IV. SYMMETRIES AND SKIN MODES

In this section, we will first use the GBZ theory to derive the emergence or the absence of skin modes in the non-Hermitian symmetry groups we concerned, namely, $G_{\mathcal{T}_\pm}$, $G_{\mathcal{T}_\pm}$, $G_{\mathcal{T}_\pm, (\mathcal{PT})_\pm}$, $G_{\mathcal{T}_\pm, (\mathcal{PT})_\pm}$, $G_{\bar{\mathcal{T}}_\pm, (\mathcal{PT})_\pm}$, $G_{\bar{\mathcal{T}}_\pm, (\mathcal{PT})_\pm}$, $G_{\mathcal{T}_\pm, (\mathcal{PC})_\pm}$, $G_{\mathcal{T}_\pm, (\mathcal{PC})_\pm}$, $G_{\bar{\mathcal{T}}_\pm, (\mathcal{PC})_\pm}$, $G_{\bar{\mathcal{T}}_\pm, (\mathcal{PC})_\pm}$. After that, we will give a numerical verification for all the symmetry groups listed above. The procedure of analytical derivation can be summarized as follows: (i) we first write down the GBZ condition based on the following Table; (ii) we then find all the symmetry related non-Bloch waves; (iii) we finally check whether the GBZ condition and the symmetry related non-Bloch waves imply $|\beta_p| = |\beta_{p+1}| = 1$ (or $|\beta_{p-1}| = |\beta_p| = |\beta_{p+1}| = |\beta_{p+2}| = 1$). According to Table. 1 in the main text, only $\bar{\mathcal{T}}_+$ and $\mathcal{P}$ implies $f(\beta, E) = f(1/\beta, E)$. Therefore, the discussion is classified as follows.

### A. Case. 1

The first case is that $G$ does not contain $\bar{\mathcal{T}}_+$ nor $\mathcal{P}$, which includes

$$G_{\mathcal{T}_\pm}, G_{\mathcal{T}_\pm, (\mathcal{PT})_\pm}, G_{\mathcal{T}_\pm, (\mathcal{PC})_\pm}, G_{\mathcal{T}_\pm, (\mathcal{PC})_\pm}. \tag{52}$$

The absence of $\bar{\mathcal{T}}_+$ and $\mathcal{P}$ implies the existence of skin modes.

TABLE II. GBZ conditions for the non-Hermitian symmetry generators [8]. For example, if $G$ contains $\mathcal{T}_-$, the GBZ conditions for the real and complex spectra are $|\beta_{p-1}| = |\beta_p| = |\beta_{p+1}| = |\beta_{p+2}|$ and $|\beta_p| = |\beta_{p+1}|$, respectively.

| $\bar{\mathcal{T}}_-$ | $E \in \mathbb{C}$ | |
|---|---|---|
| $(\mathcal{PT})_-$ | $E \in \mathbb{R}$ | $|\beta_{p-1}| = |\beta_p|$ & $|\beta_{p+1}| = |\beta_{p+2}|$ |
| $(\mathcal{P}\bar{\mathcal{C}})_-$ | $iE \in \mathbb{R}$ | |
| $(\mathcal{P}\bar{\mathcal{T}})_-$ | $E \in \mathbb{C}$ | |
| $\mathcal{T}_-$ | $E \in \mathbb{R}$ | $|\beta_{p-1}| = |\beta_p| = |\beta_{p+1}| = |\beta_{p+2}|$ |
| $\bar{\mathcal{C}}_-$ | $iE \in \mathbb{R}$ | |
| Others | $E \in \mathbb{C}$ | $|\beta_p| = |\beta_{p+1}|$ |

## B.  Case. 2

The second case is that $G$ only contains $\bar{\mathcal{T}}$, which includes

$$G_{\bar{\mathcal{T}}_\pm}, G_{\bar{\mathcal{T}}_\pm, (\mathcal{PT})_\pm}, G_{\bar{\mathcal{T}}_\pm, (\mathcal{PC})_\pm}, G_{\bar{\mathcal{T}}_\pm, (\mathcal{P}\bar{\mathcal{C}})_\pm}. \tag{53}$$

Here $G_{\bar{\mathcal{T}}_\pm}$ has been discussed in the main text. We only focus on the latter three cases.

### 1.  $G_{\bar{\mathcal{T}}_\pm, (\mathcal{PT})_\pm}$

- $G_{\bar{\mathcal{T}}_+, (\mathcal{PT})_+}$:

  According to the GBZ condition $|\beta_p| = |\beta_{p+1}|$ and the transformation of non-Bloch waves

$$
\begin{array}{ccc}
|\beta, E\rangle & \xrightarrow{\bar{\mathcal{T}}_+} & |1/\beta, E\rangle \\
{\scriptstyle (\mathcal{PT})_+} \downarrow & & \downarrow {\scriptstyle (\mathcal{PT})_+} \\
|1/\beta^*, E^*\rangle & \xrightarrow{\bar{\mathcal{T}}_+} & |\beta^*, E^*\rangle
\end{array}
$$

  one can obtain the conclusion "GBZ=BZ". To be more precise, the superposition of $|k, E\rangle$ and $|-k, E\rangle$ forms the eigenstate of the open boundary Hamiltonian with eigenenergy $E$.

- $G_{\bar{\mathcal{T}}_+, (\mathcal{PT})_-}$:

  According to the GBZ condition (i) for $E \in \mathbb{R}$, $|\beta_{p-1}| = |\beta_p|$ & $|\beta_{p+1}| = |\beta_{p+2}|$; (ii) for $E \in \mathbb{C}$, $|\beta_p| = |\beta_{p+1}|$ and the transformation of non-Bloch waves

$$
\begin{array}{ccc}
|\beta, E, \uparrow\rangle & \xrightarrow{\bar{\mathcal{T}}_+} & |1/\beta, E, \uparrow\rangle \\
{\scriptstyle (\mathcal{PT})_-} \downarrow & & \downarrow {\scriptstyle (\mathcal{PT})_-} \\
|1/\beta^*, E^*, \downarrow\rangle & \xrightarrow{\bar{\mathcal{T}}_+} & |\beta^*, E^*, \downarrow\rangle
\end{array}
$$

  one can obtain the conclusion "GBZ=BZ". To be more precise, for the nonreal eigenvalue, the eigenstate is a superposition of the following two Bloch waves, $|k, E, \uparrow\rangle$ and $|-k, E, \uparrow\rangle$; for the real eigenvalue, the open boundary eigenstate has a two fold degeneracy, which means the eigenstate is a superposition of the following four Bloch waves, $|k, E, \uparrow\rangle$, $|-k, E, \uparrow\rangle$, $|k', E, \downarrow\rangle$, $|-k', E, \downarrow\rangle$. We note that with the existence of $(\mathcal{PT})_-$ symmetry, the characteristic equation satisfies $f(\beta, E) = f(1/\beta^*, E^*)$, in which the order of $\beta$ must be a even number. This means for any given $E$ and $f(\beta, E) = 0$, there must exist even number of solutions. On the other hand, for any $E \in \mathbb{R}$, if $\beta = re^{i\theta}$ is a solution, $\beta = e^{i\theta}/r$ is also a solution, which implies the solutions must come in pairs. Specially, when $r = 1$, the solutions are also formed pairs, namely, $\beta = e^{ik}$ and $\beta = e^{ik'}$. This explains why there exist $|k', E, \downarrow\rangle$, $|-k', E, \downarrow\rangle$ in the above discussion.

- $G_{\bar{\mathcal{T}}_-, (\mathcal{PT})_+}$:

According to the GBZ condition $|\beta_{p-1}| = |\beta_p|$ & $|\beta_{p+1}| = |\beta_{p+2}|$ and the transformation of non-Bloch waves

$$
\begin{array}{ccc}
|\beta, E, \uparrow\rangle & \xrightarrow{\ \tilde{\mathcal{T}}_-\ } & |1/\beta, E, \downarrow\rangle \\
{\scriptstyle (\mathcal{PT})_+}\big\downarrow & & \big\downarrow{\scriptstyle (\mathcal{PT})_+} \\
|1/\beta^*, E^*, \uparrow\rangle & \xrightarrow{\ \bar{\mathcal{T}}_-\ } & |\beta^*, E^*, \downarrow\rangle
\end{array}
$$

one can obtain the conclusion (i) for the real spectra, "GBZ=BZ"; (ii) for the nonreal spectral, "GBZ≠BZ". This means the eigenstates of the nonreal spectra are skin modes. To be more precise, for the real eigenvalue, the open boundary eigenstate is two fold degeneracy due to $\tilde{\mathcal{T}}_-$, and a superposition of the following four Bloch waves, $|k, E, \uparrow\rangle, |-k, E, \downarrow\rangle, |k', E, \uparrow\rangle, |-k', E, \downarrow\rangle$. We further note that, due to the existence of $\mathcal{PT}$ symmetry, the system can belong to the $\mathcal{PT}$-unbroken phase, whose spectra are all reals. In this $\mathcal{PT}$-unbroken phase, it is impossible to have skin modes. However, when $\mathcal{PT}$ symmetry is spontaneously broken in the system, skin modes emerge simultaneously.

- $G_{\tilde{\mathcal{T}}_-, (\mathcal{PT})_-}$:

  According to the GBZ condition $|\beta_{p-1}| = |\beta_p|$ & $|\beta_{p+1}| = |\beta_{p+2}|$ and the transformation of non-Bloch waves

$$
\begin{array}{ccc}
|\beta, E, \uparrow\rangle & \xrightarrow{\ \tilde{\mathcal{T}}_-\ } & |1/\beta, E, \downarrow\rangle \\
{\scriptstyle (\mathcal{PT})_-}\big\downarrow & & \big\downarrow{\scriptstyle (\mathcal{PT})_-} \\
|1/\beta^*, E^*, \downarrow\rangle & \xrightarrow{\ \bar{\mathcal{T}}_-\ } & |\beta^*, E^*, \uparrow\rangle
\end{array}
$$

  one can obtain the conclusion (i) for the real spectra, "GBZ=BZ"; (ii) for the nonreal spectral, "GBZ≠BZ". This case is similar to the above case $G_{\tilde{\mathcal{T}}_-, (\mathcal{PT})_+}$.

### 2. $G_{\bar{\mathcal{T}}_\pm, (\mathcal{PC})_\pm}$

- $G_{\tilde{\mathcal{T}}_+, (\mathcal{PC})_+}$:

  According to the GBZ condition $|\beta_p| = |\beta_{p+1}|$ and the transformation of non-Bloch waves

$$
\begin{array}{ccc}
|\beta, E\rangle & \xrightarrow{\ \bar{\mathcal{T}}_+\ } & |1/\beta, E\rangle \\
{\scriptstyle (\mathcal{PC})_+}\big\downarrow & & \big\downarrow{\scriptstyle (\mathcal{PC})_+} \\
|\beta, -E\rangle & \xrightarrow{\ \tilde{\mathcal{T}}_+\ } & |1/\beta, -E\rangle
\end{array}
$$

  one can obtain the conclusion "GBZ=BZ". To be more precise, for any eigenvalue in the complex plane, the corresponding eigenstate is a superposition of $|k, E\rangle$ and $|-k, E\rangle$.

- $G_{\tilde{\mathcal{T}}_+, (\mathcal{PC})_-}$:

  According to the GBZ condition $|\beta_p| = |\beta_{p+1}|$ and the transformation of non-Bloch waves

$$
\begin{array}{ccc}
|\beta, E, \uparrow\rangle & \xrightarrow{\ \bar{\mathcal{T}}_+\ } & |1/\beta, E, \uparrow\rangle \\
{\scriptstyle (\mathcal{PC})_-}\big\downarrow & & \big\downarrow{\scriptstyle (\mathcal{PC})_-} \\
|\beta, -E, \downarrow\rangle & \xrightarrow{\ \tilde{\mathcal{T}}_+\ } & |1/\beta, -E, \downarrow\rangle
\end{array}
$$

  one can obtain the conclusion "GBZ=BZ". To be more precise, for any eigenvalue in the complex plane, the corresponding eigenstate is a superposition of $|k, E, \uparrow\rangle$ and $|-k, E, \uparrow\rangle$.

- $G_{\tilde{\mathcal{T}}_-, (\mathcal{PC})_+}$:

  According to the GBZ condition $|\beta_{p-1}| = |\beta_p|$ & $|\beta_{p+1}| = |\beta_{p+2}|$ and the transformation of non-Bloch waves

$$
\begin{array}{ccc}
|\beta, E, \uparrow\rangle & \xrightarrow{\ \tilde{\mathcal{T}}_-\ } & |1/\beta, E, \downarrow\rangle \\
{\scriptstyle (\mathcal{PC})_+}\big\downarrow & & \big\downarrow{\scriptstyle (\mathcal{PC})_+} \\
|\beta, -E, \uparrow\rangle & \xrightarrow{\ \bar{\mathcal{T}}_-\ } & |1/\beta, -E, \downarrow\rangle
\end{array}
$$

  one can obtain the conclusion "GBZ≠BZ". This means the system can have skin modes.

- $G_{\mathcal{T}_-,(\mathcal{PC})_-}$:

  According to the GBZ condition $|\beta_{p-1}| = |\beta_p|$ & $|\beta_{p+1}| = |\beta_{p+2}|$ and the transformation of non-Bloch waves

  $$
  \begin{array}{ccc}
  |\beta, E, \uparrow\rangle & \xrightarrow{\ \bar{\mathcal{T}}_-\ } & |1/\beta, E, \downarrow\rangle \\
  {\scriptstyle (\mathcal{PC})_-}\Big\downarrow & & \Big\downarrow{\scriptstyle (\mathcal{PC})_-} \\
  |\beta, -E, \downarrow\rangle & \xrightarrow{\ \bar{\mathcal{T}}_-\ } & |1/\beta, -E, \uparrow\rangle
  \end{array}
  $$

  one can obtain the conclusion "GBZ$\neq$BZ". This means the system can have skin modes.

### 3. $G_{\bar{\mathcal{T}}_\pm,(\mathcal{P}\bar{\mathcal{C}})_\pm}$

- $G_{\bar{\mathcal{T}}_+,(\mathcal{P}\bar{\mathcal{C}})_+}$:

  According to the GBZ condition $|\beta_p| = |\beta_{p+1}|$ and the transformation of non-Bloch waves

  $$
  \begin{array}{ccc}
  |\beta, E\rangle & \xrightarrow{\ \mathcal{T}_+\ } & |1/\beta, E\rangle \\
  {\scriptstyle (\mathcal{P}\bar{\mathcal{C}})_+}\Big\downarrow & & \Big\downarrow{\scriptstyle (\mathcal{P}\bar{\mathcal{C}})_+} \\
  |1/\beta^*, -E^*\rangle & \xrightarrow{\ \bar{\mathcal{T}}_+\ } & |\beta^*, -E^*\rangle
  \end{array}
  $$

  one can obtain the conclusion "GBZ=BZ". To be more precise, for any eigenvalue in the complex plane, the corresponding eigenstate is a superposition of $|k, E\rangle$ and $|-k, E\rangle$.

- $G_{\bar{\mathcal{T}}_+,(\mathcal{P}\bar{\mathcal{C}})_-}$:

  According to the GBZ condition (i) for $iE \in \mathbb{R}$, $|\beta_{p-1}| = |\beta_p|$ & $|\beta_{p+1}| = |\beta_{p+2}|$; (ii) for $E \in \mathbb{C}$, $|\beta_p| = |\beta_{p+1}|$ and the transformation of non-Bloch waves

  $$
  \begin{array}{ccc}
  |\beta, E, \uparrow\rangle & \xrightarrow{\ \bar{\mathcal{T}}_+\ } & |1/\beta, E, \uparrow\rangle \\
  {\scriptstyle (\mathcal{P}\bar{\mathcal{C}})_-}\Big\downarrow & & \Big\downarrow{\scriptstyle (\mathcal{P}\bar{\mathcal{C}})_-} \\
  |1/\beta^*, -E^*, \downarrow\rangle & \xrightarrow{\ \bar{\mathcal{T}}_+\ } & |\beta^*, -E^*, \downarrow\rangle
  \end{array}
  $$

  one can obtain the conclusion "GBZ=BZ". To be more precise, for the nonimaginary eigenvalue, the eigenstate is a superposition of $|k, E, \uparrow\rangle, |-k, E, \uparrow\rangle$; for the imaginary eigenvalue, the eigenstate is a superposition of $|k, E, \uparrow\rangle, |-k, E, \uparrow\rangle, |k', E, \downarrow\rangle, |-k', E, \downarrow\rangle$.

- $G_{\bar{\mathcal{T}}_-,(\mathcal{P}\bar{\mathcal{C}})_+}$:

  According to the GBZ condition $|\beta_{p-1}| = |\beta_p|$ & $|\beta_{p+1}| = |\beta_{p+2}|$ and the transformation of non-Bloch waves

  $$
  \begin{array}{ccc}
  |\beta, E, \uparrow\rangle & \xrightarrow{\ \bar{\mathcal{T}}_-\ } & |1/\beta, E, \downarrow\rangle \\
  {\scriptstyle (\mathcal{P}\bar{\mathcal{C}})_+}\Big\downarrow & & \Big\downarrow{\scriptstyle (\mathcal{P}\bar{\mathcal{C}})_+} \\
  |1/\beta^*, -E^*, \uparrow\rangle & \xrightarrow{\ \bar{\mathcal{T}}_-\ } & |\beta^*, -E^*, \downarrow\rangle
  \end{array}
  $$

  one can obtain the conclusion (i) for the imaginary spectra, "GBZ=BZ"; (ii) for the nonimaginary spectral, "GBZ$\neq$BZ". To be more precise, for the imaginary eigenvalue, the eigenstate is a superposition of the following four Bloch waves, $|k, E, \uparrow\rangle, |-k, E, \uparrow\rangle, |k', E, \downarrow\rangle, |-k', E, \downarrow\rangle$.

- $G_{\bar{\mathcal{T}}_-,(\mathcal{P}\bar{\mathcal{C}})_-}$:

  According to the GBZ condition $|\beta_{p-1}| = |\beta_p|$ & $|\beta_{p+1}| = |\beta_{p+2}|$ and the transformation of non-Bloch waves

  $$
  \begin{array}{ccc}
  |\beta, E, \uparrow\rangle & \xrightarrow{\ \bar{\mathcal{T}}_-\ } & |1/\beta, E, \downarrow\rangle \\
  {\scriptstyle (\mathcal{P}\bar{\mathcal{C}})_-}\Big\downarrow & & \Big\downarrow{\scriptstyle (\mathcal{P}\bar{\mathcal{C}})_-} \\
  |1/\beta^*, -E^*, \downarrow\rangle & \xrightarrow{\ \bar{\mathcal{T}}_-\ } & |\beta^*, -E^*, \uparrow\rangle
  \end{array}
  $$

  one can obtain the conclusion (i) for the imaginary spectra, "GBZ=BZ"; (ii) for the nonimaginary spectral, "GBZ$\neq$BZ". This case is similar to the above case $G_{\bar{\mathcal{T}}_-,(\mathcal{P}\bar{\mathcal{C}})_+}$.

## C. Case. 3

The third case is that $G$ only contains $\mathcal{P}$, which includes

$$G_{\mathcal{T}_\pm, (\mathcal{P}\mathcal{T})_\pm}. \tag{54}$$

- $G_{\mathcal{T}_+, (\mathcal{P}\mathcal{T})_+}$:

  According to the GBZ condition $|\beta_p| = |\beta_{p+1}|$ and the transformation of non-Bloch waves

  $$
  \begin{array}{ccc}
  |\beta, E\rangle & \xrightarrow{\mathcal{T}_+} & |\beta^*, E^*\rangle \\
  {\scriptstyle (\mathcal{P}\mathcal{T})_+}\downarrow & & \downarrow{\scriptstyle (\mathcal{P}\mathcal{T})_+} \\
  |1/\beta^*, E^*\rangle & \xrightarrow{\mathcal{T}_+} & |1/\beta, E\rangle
  \end{array}
  $$

  one can obtain the conclusion "GBZ=BZ". To be more precise, for any eigenvalue in the complex plane, the corresponding eigenstate is a superposition of $|k, E\rangle$ and $|-k, E\rangle$.

- $G_{\mathcal{T}_+, (\mathcal{P}\mathcal{T})_-}$:

  According to the GBZ condition (i) for $E \in \mathbb{R}$, $|\beta_{p-1}| = |\beta_p|$ & $|\beta_{p+1}| = |\beta_{p+2}|$; (ii) for $E \in \mathbb{C}$, $|\beta_p| = |\beta_{p+1}|$ and the transformation of non-Bloch waves

  $$
  \begin{array}{ccc}
  |\beta, E, \uparrow\rangle & \xrightarrow{\mathcal{T}_+} & |\beta^*, E^*, \uparrow\rangle \\
  {\scriptstyle (\mathcal{P}\mathcal{T})_-}\downarrow & & \downarrow{\scriptstyle (\mathcal{P}\mathcal{T})_-} \\
  |1/\beta^*, E^*, \downarrow\rangle & \xrightarrow{\mathcal{T}_+} & |1/\beta, E, \downarrow\rangle
  \end{array}
  $$

  one can obtain the conclusion "GBZ=BZ". To be more precise, for the nonreal eigenvalue, the eigenstate is a superposition of the following two Bloch waves, $|k, E, \uparrow\rangle$ and $|-k, E, \downarrow\rangle$; for the real eigenvalue, the open boundary eigenstate is a superposition of the following four Bloch waves, $|k, E, \uparrow\rangle$, $|-k, E, \downarrow\rangle$, $|k', E, \uparrow\rangle$, $|-k', E, \downarrow\rangle$. We also note that for any nonreal $E$ in the complex plane, the solution is a superposition of two spin bands. The phenomena that the open boundary eigenstate is a superposition of two distinct bands has been reported in Ref. [5].

- $G_{\mathcal{T}_-, (\mathcal{P}\mathcal{T})_+}$:

  According to the GBZ condition (i) for $E \in \mathbb{R}$, $|\beta_{p-1}| = |\beta_p| = |\beta_{p+1}| = |\beta_{p+2}|$; (ii) for $E \in \mathbb{C}$, $|\beta_p| = |\beta_{p+1}|$ and the transformation of non-Bloch waves

  $$
  \begin{array}{ccc}
  |\beta, E, \uparrow\rangle & \xrightarrow{\mathcal{T}_-} & |\beta^*, E^*, \downarrow\rangle \\
  {\scriptstyle (\mathcal{P}\mathcal{T})_+}\downarrow & & \downarrow{\scriptstyle (\mathcal{P}\mathcal{T})_+} \\
  |1/\beta^*, E^*, \uparrow\rangle & \xrightarrow{\mathcal{T}_-} & |1/\beta, E, \downarrow\rangle
  \end{array}
  $$

  one can obtain the conclusion "GBZ=BZ". This case is similar to the above case $G_{\mathcal{T}_+, (\mathcal{P}\mathcal{T})_-}$.

- $G_{\mathcal{T}_-, (\mathcal{P}\mathcal{T})_-}$:

  According to the GBZ condition (i) for $E \in \mathbb{R}$, $|\beta_{p-1}| = |\beta_p| = |\beta_{p+1}| = |\beta_{p+2}|$; (ii) for $E \in \mathbb{C}$, $|\beta_p| = |\beta_{p+1}|$ and the transformation of non-Bloch waves

  $$
  \begin{array}{ccc}
  |\beta, E, \uparrow\rangle & \xrightarrow{\mathcal{T}_-} & |\beta^*, E^*, \downarrow\rangle \\
  {\scriptstyle (\mathcal{P}\mathcal{T})_-}\downarrow & & \downarrow{\scriptstyle (\mathcal{P}\mathcal{T})_-} \\
  |1/\beta^*, E^*, \downarrow\rangle & \xrightarrow{\mathcal{T}_-} & |1/\beta, E, \uparrow\rangle
  \end{array}
  $$

  one can obtain the conclusion "GBZ=BZ". To be more precise, for the nonreal eigenvalue, the eigenstate is a superposition of the following two Bloch waves, $|k, E, \uparrow\rangle$ and $|-k, E, \uparrow\rangle$; for the real eigenvalue, the open boundary eigenstate is a superposition of the following four Bloch waves, $|k, E, \uparrow\rangle$, $|-k, E, \uparrow\rangle$, $|k', E, \downarrow\rangle$, $|-k', E, \downarrow\rangle$.

### D. Case. 4

The last case is that $G$ contains both $\bar{\mathcal{T}}_+$ and $\mathcal{P}$, which includes

$$G_{\bar{\mathcal{T}}_\pm,(\mathcal{P}\bar{\mathcal{T}})_\pm}.\tag{55}$$

- $G_{\bar{\mathcal{T}}_+,(\mathcal{P}\bar{\mathcal{T}})_+}$:

  According to the GBZ condition $|\beta_p| = |\beta_{p+1}|$ and the transformation of non-Bloch waves

$$
\begin{array}{ccc}
|\beta, E\rangle & \xrightarrow{\bar{\mathcal{T}}_+} & |1/\beta, E\rangle \\
{\scriptstyle (\mathcal{P}\mathcal{T})_+}\downarrow & & \downarrow{\scriptstyle (\mathcal{P}\mathcal{T})_+} \\
|\beta, E\rangle & \xrightarrow{\bar{\mathcal{T}}_+} & |1/\beta, E\rangle
\end{array}
$$

  one can obtain the conclusion "GBZ=BZ". To be more precise, for any eigenvalue in the complex plane, the corresponding eigenstate is a superposition of $|k, E\rangle$ and $|-k, E\rangle$.

- $G_{\bar{\mathcal{T}}_+,(\mathcal{P}\bar{\mathcal{T}})_-}$:

  According to the GBZ condition $|\beta_{p-1}| = |\beta_p| = |\beta_{p+1}| = |\beta_{p+2}|$ and the transformation of non-Bloch waves

$$
\begin{array}{ccc}
|\beta, E, \uparrow\rangle & \xrightarrow{\bar{\mathcal{T}}_+} & |1/\beta, E, \uparrow\rangle \\
{\scriptstyle (\mathcal{P}\mathcal{T})_-}\downarrow & & \downarrow{\scriptstyle (\mathcal{P}\mathcal{T})_-} \\
|\beta, E, \downarrow\rangle & \xrightarrow{\bar{\mathcal{T}}_+} & |1/\beta, E, \downarrow\rangle
\end{array}
$$

  one can obtain the conclusion "GBZ=BZ". To be more precise, for any eigenvalue in the complex plane, the corresponding eigenstate is a superposition of $|k, E, \uparrow\rangle, |-k, E, \uparrow\rangle, |k, E, \downarrow\rangle, |-k, E, \downarrow\rangle$.

- $G_{\bar{\mathcal{T}}_-,(\mathcal{P}\bar{\mathcal{T}})_+}$:

  According to the GBZ condition $|\beta_{p-1}| = |\beta_p|$ & $|\beta_{p+1}| = |\beta_{p+2}|$ and the transformation of non-Bloch waves

$$
\begin{array}{ccc}
|\beta, E, \uparrow\rangle & \xrightarrow{\bar{\mathcal{T}}_-} & |1/\beta, E, \downarrow\rangle \\
{\scriptstyle (\mathcal{P}\bar{\mathcal{T}})_+}\downarrow & & \downarrow{\scriptstyle (\mathcal{P}\bar{\mathcal{T}})_+} \\
|\beta, E, \uparrow\rangle & \xrightarrow{\bar{\mathcal{T}}_-} & |1/\beta, E, \downarrow\rangle
\end{array}
$$

  one can obtain the conclusion "GBZ$\neq$BZ".

- $G_{\bar{\mathcal{T}}_-,(\mathcal{P}\bar{\mathcal{T}})_-}$:

  According to the GBZ condition $|\beta_{p-1}| = |\beta_p| = |\beta_{p+1}| = |\beta_{p+2}|$ and the transformation of non-Bloch waves

$$
\begin{array}{ccc}
|\beta, E, \uparrow\rangle & \xrightarrow{\bar{\mathcal{T}}_-} & |1/\beta, E, \downarrow\rangle \\
{\scriptstyle (\mathcal{P}\mathcal{T})_-}\downarrow & & \downarrow{\scriptstyle (\mathcal{P}\mathcal{T})_-} \\
|\beta, E, \downarrow\rangle & \xrightarrow{\bar{\mathcal{T}}_-} & |1/\beta, E, \uparrow\rangle
\end{array}
$$

  one can obtain the conclusion "GBZ=BZ". To be more precise, for any eigenvalue in the complex plane, the corresponding eigenstate is a superposition of $|k, E, \uparrow\rangle, |-k, E, \uparrow\rangle, |k, E, \downarrow\rangle, |-k, E, \downarrow\rangle$.

TABLE III. Non-Hermitian symmetry classes we concerned

| Group generators | $\mathcal{T}$ | $\bar{\mathcal{T}}$ |
|---|---|---|
| $U_X^{-1}\mathcal{H}(k)U_X =$ | $\mathcal{H}^*(-k)$ | $\mathcal{H}^t(-k)$ |
| Representations | $\mathcal{T} = \mathcal{K}^*$ | $\bar{\mathcal{T}} = \mathcal{K}^t$ |
| | $\mathcal{T} = s_y\mathcal{K}^*$ | $\bar{\mathcal{T}} = s_y\mathcal{K}^t$ |

In summary, skin modes are absent when $G$ contains $\bar{\mathcal{T}}_+$, or $\mathcal{P}$ with an exceptional case, namely, $G_{\bar{\mathcal{T}}_-,(\mathcal{P}\bar{\mathcal{T}})_+}$, where the skin modes emerge with the presence of IS.

### E. Numerical verification

TABLE IV. Non-Hermitian symmetry classes we concerned.

| Group generators | $(\mathcal{T},\mathcal{PT})$ | $(\bar{\mathcal{T}},\mathcal{P}\bar{\mathcal{T}})$ | $(\mathcal{T},\mathcal{P}\bar{\mathcal{T}})$ | $(\bar{\mathcal{T}},\mathcal{PT})$ |
|---|---|---|---|---|
| $U_X^{-1}\mathcal{H}(k)U_X =$ | $(\mathcal{H}^*(-k),\mathcal{H}^*(k))$ | $(\mathcal{H}^t(-k),\mathcal{H}^t(k))$ | $(\mathcal{H}^*(-k),\mathcal{H}^t(k))$ | $(\mathcal{H}^t(-k),\mathcal{H}^*(k))$ |
| Representations | $\mathcal{T}=\mathcal{K}^*, \mathcal{PT}=\sigma_x\mathcal{K}^*$ | $\bar{\mathcal{T}}=\mathcal{K}^t, \mathcal{P}\bar{\mathcal{T}}=\sigma_x\mathcal{K}^t$ | $\mathcal{T}=\mathcal{K}^*, \mathcal{P}\bar{\mathcal{T}}=\sigma_x\mathcal{K}^t$ | $\bar{\mathcal{T}}=\mathcal{K}^t, \mathcal{PT}=\sigma_x\mathcal{K}^*$ |
| | $\mathcal{T}=\mathcal{K}^*, \mathcal{PT}=\sigma_y\mathcal{K}^*$ | $\bar{\mathcal{T}}=\mathcal{K}^t, \mathcal{P}\bar{\mathcal{T}}=\sigma_y\mathcal{K}^t$ | $\mathcal{T}=\mathcal{K}^*, \mathcal{P}\bar{\mathcal{T}}=\sigma_y\mathcal{K}^t$ | $\bar{\mathcal{T}}=\mathcal{K}^t, \mathcal{PT}=\sigma_y\mathcal{K}^*$ |
| | $\mathcal{T}=s_y\mathcal{K}^*, \mathcal{PT}=\sigma_x s_y\mathcal{K}^*$ | $\bar{\mathcal{T}}=s_y\mathcal{K}^t, \mathcal{P}\bar{\mathcal{T}}=\sigma_x s_y\mathcal{K}^t$ | $\mathcal{T}=s_y\mathcal{K}^*, \mathcal{P}\bar{\mathcal{T}}=\sigma_x s_y\mathcal{K}^t$ | $\bar{\mathcal{T}}=s_y\mathcal{K}^t, \mathcal{PT}=\sigma_x s_y\mathcal{K}^*$ |
| | $\mathcal{T}=s_y\mathcal{K}^*, \mathcal{PT}=\sigma_y s_y\mathcal{K}^*$ | $\bar{\mathcal{T}}=s_y\mathcal{K}^t, \mathcal{P}\bar{\mathcal{T}}=\sigma_y s_y\mathcal{K}^t$ | $\mathcal{T}=s_y\mathcal{K}^*, \mathcal{P}\bar{\mathcal{T}}=\sigma_y s_y\mathcal{K}^t$ | $\bar{\mathcal{T}}=s_y\mathcal{K}^t, \mathcal{PT}=\sigma_y s_y\mathcal{K}^*$ |
| Group generators | $\mathcal{T},\mathcal{PC}$ | $(\mathcal{T},\mathcal{P}\bar{\mathcal{C}})$ | $(\bar{\mathcal{T}},\mathcal{PC})$ | $(\bar{\mathcal{T}},\mathcal{P}\bar{\mathcal{C}})$ |
| $U_X^{-1}\mathcal{H}(k)U_X =$ | $(\mathcal{H}^*(-k),-\mathcal{H}^t(k))$ | $(\mathcal{H}^*(-k),-\mathcal{H}^*(k))$ | $(\mathcal{H}^t(-k),-\mathcal{H}^t(k))$ | $(\mathcal{H}^t(-k),-\mathcal{H}^*(k))$ |
| Representations | $\mathcal{T}=\mathcal{K}^*, \mathcal{PC}=\sigma_x\mathcal{K}^t$ | $\mathcal{T}=\mathcal{K}^*, \mathcal{P}\bar{\mathcal{C}}=\sigma_x\mathcal{K}^*$ | $\bar{\mathcal{T}}=\mathcal{K}^t, \mathcal{PC}=\sigma_x\mathcal{K}^t$ | $\bar{\mathcal{T}}=\mathcal{K}^t, \mathcal{P}\bar{\mathcal{C}}=\sigma_x\mathcal{K}^*$ |
| | $\mathcal{T}=\mathcal{K}^*, \mathcal{PC}=\sigma_y\mathcal{K}^t$ | $\mathcal{T}=\mathcal{K}^*, \mathcal{P}\bar{\mathcal{C}}=\sigma_y\mathcal{K}^*$ | $\bar{\mathcal{T}}=\mathcal{K}^t, \mathcal{PC}=\sigma_y\mathcal{K}^t$ | $\bar{\mathcal{T}}=\mathcal{K}^t, \mathcal{P}\bar{\mathcal{C}}=\sigma_y\mathcal{K}^*$ |
| | $\mathcal{T}=s_y\mathcal{K}^*, \mathcal{PC}=\sigma_x s_y\mathcal{K}^t$ | $\mathcal{T}=s_y\mathcal{K}^*, \mathcal{P}\bar{\mathcal{C}}=\sigma_x s_y\mathcal{K}^*$ | $\bar{\mathcal{T}}=s_y\mathcal{K}^t, \mathcal{PC}=\sigma_x s_y\mathcal{K}^t$ | $\bar{\mathcal{T}}=s_y\mathcal{K}^t, \mathcal{P}\bar{\mathcal{C}}=\sigma_x s_y\mathcal{K}^*$ |
| | $\mathcal{T}=s_y\mathcal{K}^*, \mathcal{PC}=\sigma_y s_y\mathcal{K}^t$ | $\mathcal{T}=s_y\mathcal{K}^*, \mathcal{P}\bar{\mathcal{C}}=\sigma_y s_y\mathcal{K}^*$ | $\bar{\mathcal{T}}=s_y\mathcal{K}^t, \mathcal{PC}=\sigma_y s_y\mathcal{K}^t$ | $\bar{\mathcal{T}}=s_y\mathcal{K}^t, \mathcal{P}\bar{\mathcal{C}}=\sigma_y s_y\mathcal{K}^*$ |

In order to verify the above conclusion, we provide a complete numerical calculation for all symmetry groups we concerned. We consider a four-band Bloch Hamiltonian

$$\mathcal{H}(k) = \sum_{\mu\nu} h_{\mu\nu}(k)\sigma_\mu \otimes \tau_\nu. \tag{56}$$

The representations of the symmetries are shown in Table. III and IV. Under the restriction of these symmetries, $h_{\mu\nu}(k)$ can be even or odd functions. In the numerical calculation, the even/odd functions are chosen to be

$$h_{\mu\nu}^{\text{even}}(k) = a_{\mu\nu} + b_{\mu\nu}\cos k + c_{\mu\nu}\cos^2 k, \qquad h_{\mu\nu}^{\text{odd}}(k) = e_{\mu\nu}\sin k + f_{\mu\nu}\sin^3 k. \tag{57}$$

For the real parameters, they are chosen to be random numbers in the region $[-10, 10]$; for the complex parameters, both the real and the imaginary parts are chosen to be random numbers in the region $[-10, 10]$. The existence of skin modes can be indicated by the discrepancy between periodic/open boundary spectrum shown in Fig. 1 and Fig. 2 with gray/black points [4].

## V. NON-HERMITIAN RICE-MELE MODEL

In order to make the Supplemental Materials self-contained, we start from the conventional Rice-Mele model which can be expressed as:

$$\mathcal{H}_{\text{RM}}(k) = (t_1 + t_2\cos k)\sigma_x + t_2\sin k\sigma_y + \mu\sigma_z. \tag{58}$$

When $\mu = 0$, the model reduces to the Su-Schrieffer-Heeger model.

### A. The condition for the emergence of skin modes

In this subsection, we will discuss how to induce skin modes in the Rice-Mele model. Here the non-Hermitian terms are not restricted to be on-site dissipations as shown in Eq. 2 in the main text. Now consider the following Hamiltonian,

$$\mathcal{H}_{\text{nRM}}(k) = \mathcal{H}_{\text{RM}}(k) + \mathcal{H}'(k), \tag{59}$$

where $\mathcal{H}'(k)$ can have both Hermitian and non-Hermitian terms.

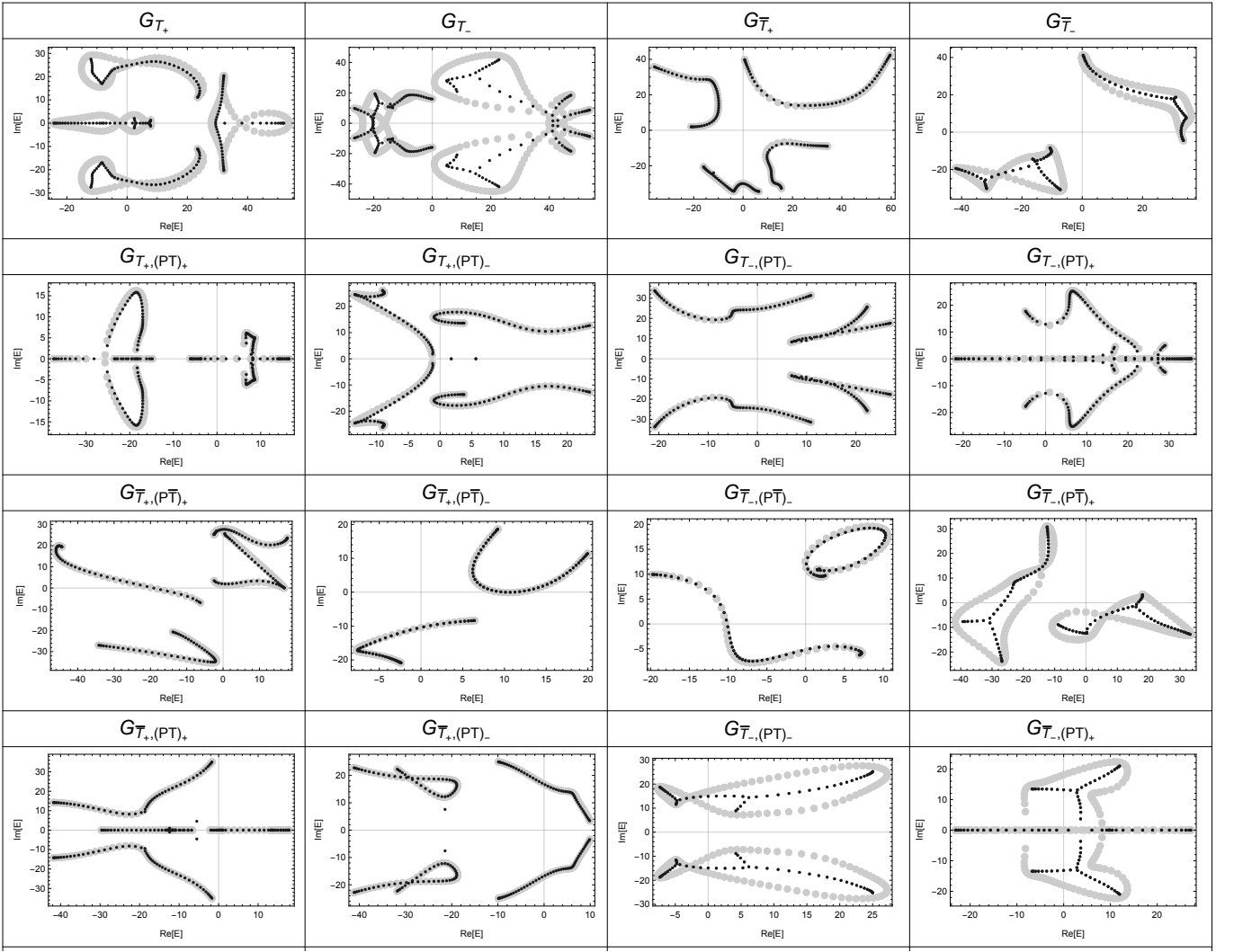

FIG. 1. Numerical verification for the emergence/absence of skin modes. The existence of skin modes can be indicated by the discrepancy between periodic/open boundary spectrum with gray/black points [4].

### 1. Spinless case

We first focus on the spinless case. It can be verified that the Rice-Mele model preserves the following symmetries,

$$\mathcal{T}_+ = \mathcal{K}^*, \quad (\mathcal{PC})_- = \sigma_y \mathcal{K}^t, \quad \mathcal{PCT} = \sigma_y \mathcal{K}^\dagger. \tag{60}$$

According to the discussion in above section, skin modes are absent when $G$ contains $\bar{\mathcal{T}}_+$, or $\mathcal{P}$ , except the case $G_{\bar{\mathcal{T}}_-,(\mathcal{P}\bar{\mathcal{T}})_+}$, where the skin modes emerge with the presence of IS. Since the Rice-Mele model breaks the IS, in order to have skin modes, spinless TRS$^\dagger$ ($\mathcal{T}_+$) must also be broken in $\mathcal{H}_{\mathrm{nRM}}(k)$. This implies

$$\mathcal{H}_{\mathrm{RM}}^t(k) + \mathcal{H}'^t(k) \neq \mathcal{H}_{\mathrm{RM}}(-k) + \mathcal{H}'(-k) \; \rightarrow \; \mathcal{H}'^t(k) \neq \mathcal{H}'(-k). \tag{61}$$

The solution of the above equation is

$$\mathcal{H}'(k) \neq \sum_\mu h_\mu(k)\sigma_\mu, \tag{62}$$

where $\mu = 0, x, y, z$ and

$$h_{0,x,z}(k) = h_{0,x,z}(-k), \qquad h_y(k) = -h_y(-k). \tag{63}$$

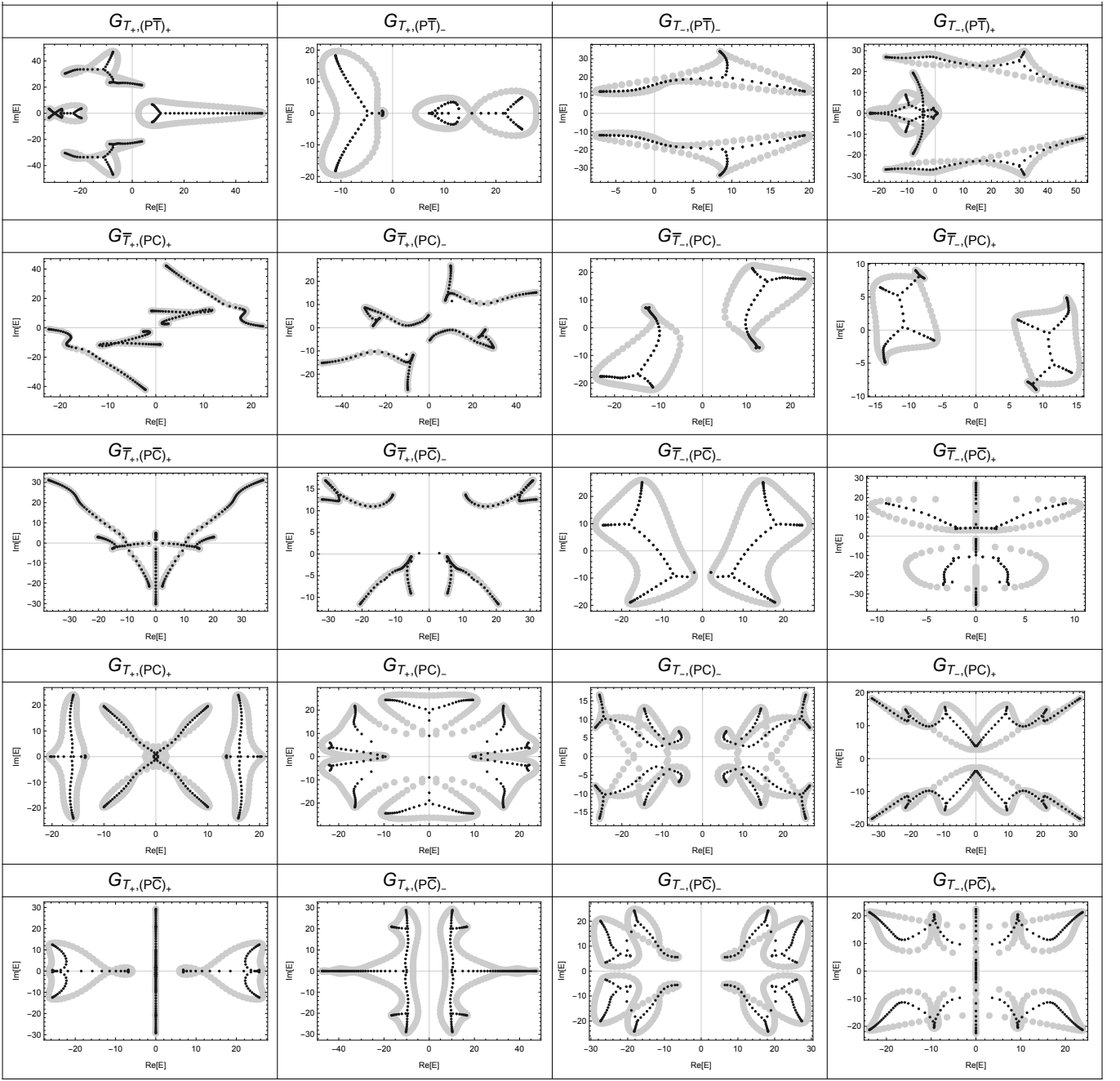

FIG. 2. Numerical verification for the emergence/absence of skin modes. The existence of skin modes can be indicated by the discrepancy between periodic/open boundary spectrum with gray/black points [4].

Based on the above equations, we can list some possible terms of $\mathcal{H}'(k)$, e.g.,

$$
\begin{aligned}
&1. \ \mathcal{H}'(k) = ig_0(k)\sigma_0, && g_0(k) = -g_0(-k), \\
&2. \ \mathcal{H}'(k) = ig_z(k)\sigma_z, && g_z(k) = -g_z(-k), \\
&3. \ \mathcal{H}'(k) = [h_z(k) + ig_z(k)]\sigma_z, && h_z(k) = -h_z(-k), \quad g_z(k) = g_z(-k).
\end{aligned}
\tag{64}
$$

The last case with $h_z(k) = \lambda \sin k$ and $g_z(k) = \gamma$ is the spinless model discussed in the main text. Here we note that if $\mathcal{H}_{\mathrm{nRM}}(k)$ is restricted to the form of Eq. 2 in the main text, the skin modes can only be induced when TRS is broken in the Hermitian part of $\mathcal{H}_{\mathrm{nRM}}(k)$. This is because $\mathcal{H}_{\mathrm{nRM}}(k)$ always have $\bar{\mathcal{T}}_+$ due to $[U_{\mathcal{T}_+}, \Gamma_0] = 0$.

### 2. Spinful case

If we add spin-orbit couping to the Hermitian Rice-Mele model, the TRS becomes $\mathcal{T}_- = is_y\mathcal{K}^*$, where $s_\mu$ represents the spin Pauli matrix. Since the spinful Rice-Mele model breaks both $\mathcal{T}_+$ and $\mathcal{P}$, the corresponding $\mathcal{H}_{\mathrm{nRM}}(k)$ can no longer have $\bar{\mathcal{T}}_+$ and $\mathcal{P}$ any more. Therefore, it is possible to have skin modes by adding arbitrary kind of non-Hermitian term.

## B. Phase diagram of spinless model

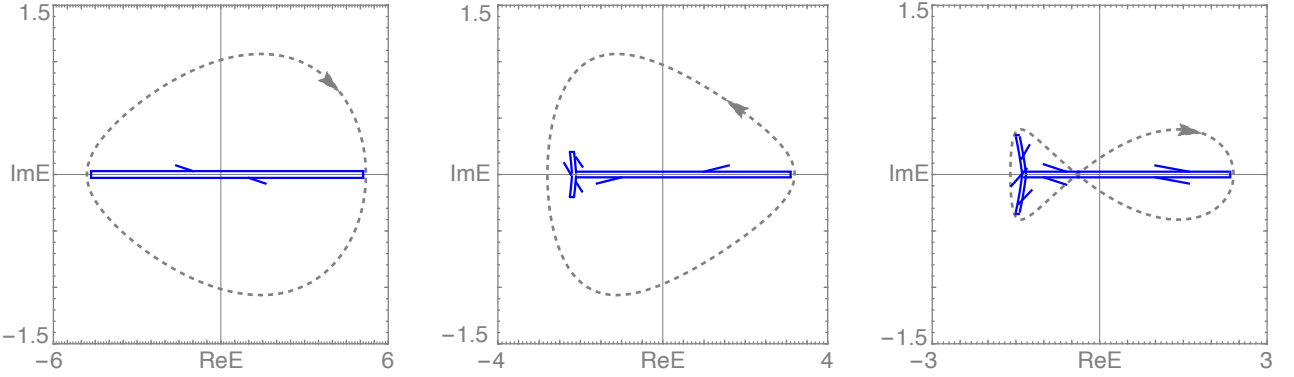

FIG. 3. Several examples of the zero and nonzero area curves. The blue ones represent the nonzero area cures. The arrows shows the moving direction as $k$ evolves. One can notice that any point on the cure must be covered twice with opposite moving directions.

In the main text, we mentioned that regardless the value of $\mu$, the skin modes always exist when $\lambda t_1 t_2 \neq 0$. Now we show this. The Bloch Hamiltonian of the spinless model we proposed in the main text has the following form

$$\mathcal{H}_{\mathrm{spinless}}(k) = \mathcal{H}_{\mathrm{RM}}(k) + \lambda \sin k\sigma_z + i\gamma\sigma_z, \tag{65}$$

where $\lambda$ is the strength of $\pi$-flux, which breaks the TRS when $\gamma = 0$. The existence of skin modes can be predicted by the winding number [4, 9]

$$\nu(E_0) = \frac{1}{2\pi i}\int_0^{2\pi} dk\partial_k \ln\det[\mathcal{H}_{\mathrm{spinless}}(k) - E_0]. \tag{66}$$

It should be noted that the above winding number formula depends on the choice of $E_0$. The statement of the existence of skin modes is claimed as follow:

*If there exist a $E_0 \in \mathbb{C}$ such that the winding number is nonzero, then, there must exist skin modes.*

An equivalent statement is that [4]

*If the area of the parametric curve defined as follow*

$$k \quad \to \quad (\mathrm{Re}[\det[\mathcal{H}_{\mathrm{spinless}}(k)]], \mathrm{Im}[\det[\mathcal{H}_{\mathrm{spinless}}(k)]]) \tag{67}$$

*is nonzero, then, there must exist skin modes.*

The reason is that if the area is nonzero, one can always find a point $E_0$ in the region contributing to the area, such that the winding number is nonzero. Since the second statement is easy to apply, we will use it to calculate the phase diagram. The determinant of $\mathcal{H}_{\mathrm{spinless}}(k)$ can be expressed as

$$\det[\mathcal{H}_{\mathrm{spinless}}(k)] = \frac{\lambda^2}{2}\cos 2k - 2\lambda\mu\sin k - 2t_1 t_2\cos k + (\gamma^2 - t_1^2 - t_2^2 - \mu^2 - \frac{\lambda^2}{2}) - i(2\lambda\gamma\sin k + 2\gamma\mu). \tag{68}$$

In order to have a zero area, any point of $(\mathrm{Re}[\det[\mathcal{H}_{\mathrm{spinless}}(k)]], \mathrm{Im}[\det[\mathcal{H}_{\mathrm{spinless}}(k)]])$ in the complex plane must be covered twice with opposite moving directions as $k$ evolves, as shown in Fig. 3.

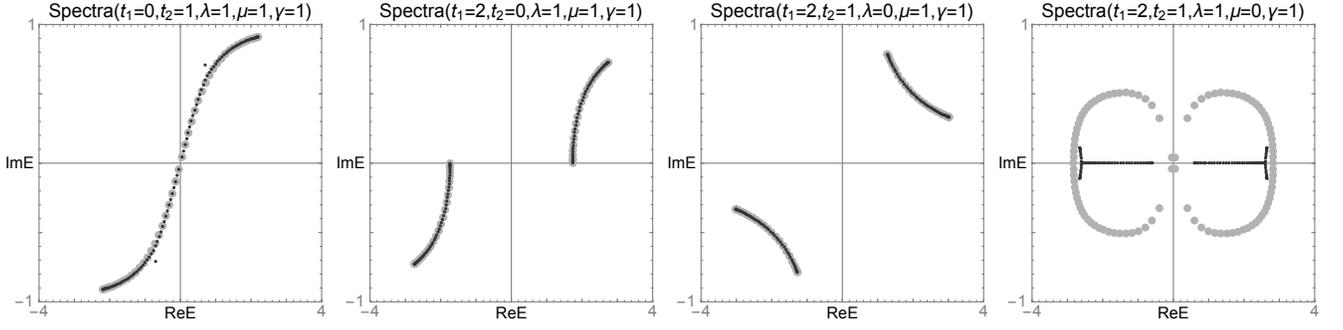

FIG. 4. Comparison between the open/periodic boundary spectra of the spinless model Eq. 65 with different parameters. The gray/black points represent the periodic/open boundary spectra.

- If the imaginary part is $k$ independent, the area must be zero. In this case, we require

$$\lambda\gamma = 0. \tag{69}$$

Since $\gamma$ is the only non-Hermitian term we added, it can not be zero. Therefore, in order to satisfy the condition, $\lambda$ must be zero. This means the preservation of TRS.

- If $\lambda\gamma \neq 0$, the following values of $k$ correspond to the same imaginary part, namely, $k$ and $\pm(2n+1)\pi - k$. Putting these values into the real part, according to $\cos 2k = \cos 2(\pm(2n+1)\pi - k)$, $\cos k \neq \cos(\pm(2n+1)\pi - k)$, one can notice that when

$$t_1 t_2 \neq 0, \tag{70}$$

$k$ and $\pm(2n+1)\pi - k$ do not correspond to the same real parts. This means the area of the curve can not be zero, since the mapping from $k \in [-\pi, \pi]$ to $\det[\mathcal{H}_{\text{spinless}}(k)]$ is a one-to-one mapping.

- If $\lambda\gamma \neq 0$ but $t_1 t_2 = 0$, the area must be zero, since $\partial_k \det[\mathcal{H}_{\text{spinless}}(k)]|_{k_0} \neq \partial_k \det[\mathcal{H}_{\text{spinless}}(k)]|_{\pm(2n+1)\pi - k_0}$. This means the moving direction of the two points $k_0$ and $\pm(2n+1)\pi - k_0$ are not the same. The contribution to the winding number is canceled.

In summary, the skin modes always exist with when $\lambda t_1 t_2 \neq 0$. As shown in Fig. 4, the numerical calculation also supports our analysis. Especially, when $\mu = 0$, as the right one of Fig. 4, on the one hand, the Rice-Mele model reduces to SSH model; on the other hand, the system also has skin modes. Indeed, the above analysis shows that the emergence of skin modes does not depend on the value of $\mu$. This means the time-reversal-breaking SSH model can also have skin modes.

## C. Spin $U(1)$ symmetry of the spinful model

The Bloch Hamiltonian of the spinful model is

$$
\begin{aligned}
\mathcal{H}_{\text{spinful}}(k) &= \mathcal{H}_{\text{RM}}(k)s_0 + \mathcal{H}_{\text{soc}}(k) + i\gamma\sigma_z s_0, \\
\mathcal{H}_{\text{soc}}(k) &= \lambda_I \sin k\sigma_z s_z - \lambda_R \sigma_y\left(s_x - \sqrt{3}s_y\right)/2.
\end{aligned}
\tag{71}
$$

When $\lambda_R = 0$, the above model also has spin $U(1)$ symmetry. This mean the Bloch Hamiltonian can be reduced to two spin blocks,

$$\mathcal{H}_{\uparrow/\downarrow}(k) = \mathcal{H}_{\text{RM}}(k) \pm \lambda_I \sin k\sigma_z + i\gamma\sigma_z. \tag{72}$$

Under the following parameters, $t_1 = \lambda_I = 2, t_2 = \mu = \gamma = 1$, the spin up block Hamiltonian reduces to the spinless model discussed in the main text, namely, Fig. 1 in the main text. For the spin down block, we show the corresponding spectra, GBZ and auxiliary GBZ in Fig. 5. One can notice the GBZ is larger than 1. The Mathematica code for the calculation of auxiliary GBZ with nonzero $\lambda_R$ is shown in the last section of the Supplemental Materials.

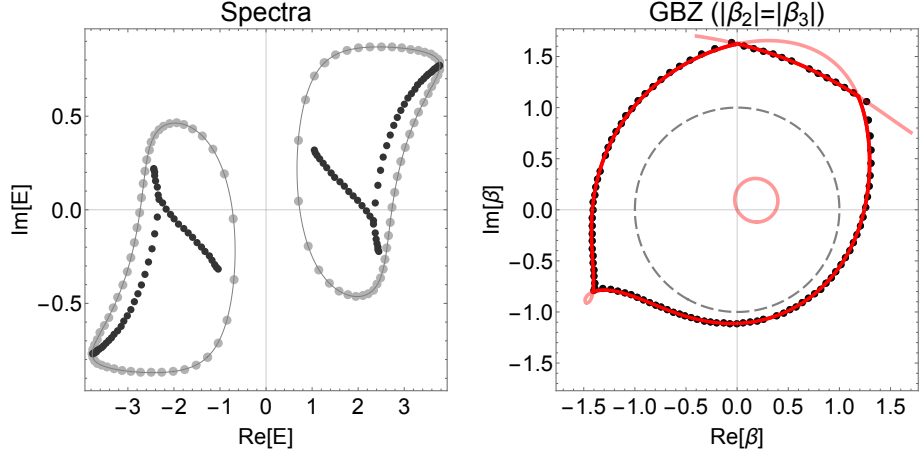

FIG. 5. (a) shows the open (black points) and periodic (gray points) boundary spectra of $\mathcal{H}_\downarrow(k)$ with the following parameters $t_1 = \lambda_I = 2, t_2 = \mu = \gamma = 1$. (b) shows the corresponding GBZ (black points) and auxiliary GBZ (red lines).

## VI. LOCAL DENSITY OF STATES AND CHIRAL TUNNELING EFFECT

In this section, we will derive Eq. 6 and Eq. 7 in the main text.

### A. Reviews of bi-orthogonal basis

We first review the bi-orthogonal quantum mechanics [10], which will be used in the following discussion. Consider the following non-Hermitian Hamiltonian

$$\mathcal{H}_{eff} = \mathcal{H} + i\Gamma, \qquad \mathcal{H}_{eff}^\dagger = \mathcal{H} - i\Gamma, \tag{73}$$

with the following eigenequations

$$
\begin{aligned}
\mathcal{H}_{eff}|\psi_n^R\rangle = E_n|\psi_n^R\rangle, \qquad &\langle\psi_n^R|\mathcal{H}_{eff}^\dagger = E_n^*\langle\psi_n^R|, \\
\mathcal{H}_{eff}^\dagger|\psi_n^L\rangle = E_n^*|\psi_n^L\rangle, \qquad &\langle\psi_n^L|\mathcal{H}_{eff} = E_n\langle\psi_n^L|.
\end{aligned}
\tag{74}
$$

We assume there is no degeneracy in the following discussion. Now consider two general right eigenstates $|\psi_n^R\rangle$ and $|\psi_m^R\rangle$. According to $2\mathcal{H} = \mathcal{H}_{eff} + \mathcal{H}_{eff}^\dagger$ and $2i\Gamma = \mathcal{H}_{eff} - \mathcal{H}_{eff}^\dagger$, we have

$$\langle\psi_m^R|2\mathcal{H}|\psi_n^R\rangle = \langle\psi_m^R|(\mathcal{H}_{eff} + \mathcal{H}_{eff}^\dagger)|\psi_n^R\rangle = (E_n + E_m^*)\langle\psi_m^R|\psi_n^R\rangle, \tag{75}$$

which in general results

$$\langle\psi_m^R|\psi_n^R\rangle = 2\frac{\langle\psi_m^R|\mathcal{H}|\psi_n^R\rangle}{E_n + E_m^*} = 2i\frac{\langle\psi_m^R|\Gamma|\psi_n^R\rangle}{E_n - E_m^*} \neq 0. \tag{76}$$

This means, in general, the right eigenstates do not satisfy the orthogonal relation. On the other hand,

$$\langle\psi_m^L|\mathcal{H}_{eff}|\psi_n^R\rangle = E_n\langle\psi_m^L|\psi_n^R\rangle = E_m\langle\psi_m^L|\psi_n^R\rangle. \tag{77}$$

Thus if $E_m \neq E_n$, $\langle\psi_m^L|\psi_n^R\rangle$ must be zero. As a result, we obtain the following identity operator [10]

$$\sum_n \frac{|\psi_n^R\rangle\langle\psi_n^L|}{\langle\psi_n^L|\psi_n^R\rangle} = \mathbb{1}, \qquad \sum_n \frac{|\psi_n^L\rangle\langle\psi_n^R|}{\langle\psi_n^R|\psi_n^L\rangle} = \mathbb{1}. \tag{78}$$

Without loss of generality, we can choose the following bi-orthogonal normalization condition in the following discussion

$$\langle\psi_m^L|\psi_n^R\rangle = \delta_{mn}. \tag{79}$$

Under this condition, the Hamiltonian can be expressed as

$$\mathcal{H}_{eff} = \sum_n E_n |\psi_n^R\rangle\langle\psi_n^L|, \qquad \mathcal{H}_{eff}^\dagger = \sum_n E_n^* |\psi_n^L\rangle\langle\psi_n^R|; \tag{80}$$

the time-evolution operator can be expressed as

$$U_{eff}(t) = e^{-i\mathcal{H}_{eff}t} = \sum_n e^{-iE_n t}|\psi_n^R\rangle\langle\psi_n^L|, \qquad U_{eff}^\dagger(t) = e^{i\mathcal{H}_{eff}^\dagger t} = \sum_n e^{iE_n^* t}|\psi_n^L\rangle\langle\psi_n^R|; \tag{81}$$

and the corresponding retarded Green's function can be expressed as

$$G(\omega) = \frac{1}{\omega - \mathcal{H}_{eff} + i\eta} = \sum_n \frac{|\psi_n^R\rangle\langle\psi_n^L|}{\omega - E_n + i\eta}, \tag{82}$$

where $\eta = 0^+$.

## B. The derivation of Eq. 6 in the main text

Although the model discussed in the main text is the Bloch Hamiltonian $\mathcal{H}(k) := \mathcal{H}_{\text{spinless}}(k) - i\gamma\sigma_0$ with open boundary condition, i.e., $H_{OBC}$, our derivation of Eq. 6 and Eq. 7 can be applied to general non-Hermitian Hamiltonians with open boundary. The retarded Green's function of $H_{OBC}$ is defined as

$$G^R(\omega) = \frac{1}{\omega - H_{OBC} + i\eta}, \tag{83}$$

where $\eta = 0^+$. Based on Eq. 82, the local DoS of $H_{OBC}$ at site $i$ is defined as

$$\nu_i(\omega) = -\frac{1}{\pi}\,\text{Im}[\langle i|G^R(\omega)|i\rangle] = -\frac{1}{\pi}\,\text{Im}\left[\sum_n \frac{\langle i|\psi_n^R\rangle\langle\psi_n^L|i\rangle}{\omega - E_n + i\eta}\right], \tag{84}$$

where $|\psi_n^{R/L}\rangle$ are the right/left eigenstates of $H_{OBC}$,

$$H_{OBC}|\psi_n^R\rangle = E_n|\psi_n^R\rangle, \quad H_{OBC}^\dagger|\psi_n^L\rangle = E_n^*|\psi_n^L\rangle. \tag{85}$$

### 1. Right eigenstates

It has been shown that in the thermodynamic limit, the asymptotic right eigenstate of $H_{OBC}$, i.e., $|\psi_n^R\rangle$ (whose eigenvalue is $E_n$) is a superposition of two generalized Bloch waves [3–5],

$$|\beta_n^R\rangle := |\psi_n^R\rangle = c_{n,p}|\beta_{n,p}^R\rangle + c_{n,p+1}|\beta_{n,p+1}^R\rangle, \tag{86}$$

where $\beta_{n,\mu}^R$ is the $\mu$th largest root (ordered by the absolute value) of $f(\beta, E_n) = 0$ and $p$ is the order of the pole of $f(\beta, E_n) = 0$. Here $f(\beta, E) = \det[E - \mathcal{H}(\beta)]$ is the characteristic equation of the non-Bloch Hamiltonian $\mathcal{H}(\beta)$ related to $H_{OBC}$. In real space, $|\beta_{n,\mu}^R\rangle$ can be expressed as

$$\begin{aligned}
\langle r|\beta_{n,\mu}^R\rangle &= (\beta_{n,\mu}^R, ..., \beta_{n,\mu}^{R,N})^t = (r_n e^{ik_{n,\mu}^R} ..., r_n^N e^{ik_{n,\mu}^R N})^t \\
&= U(r_n)(e^{ik_{n,\mu}^R}, ..., e^{ik_{n,\mu}^R N})^t \\
&= U(r_n)\langle r|k_{n,\mu}^R\rangle,
\end{aligned} \tag{87}$$

where $U(r_n) = \text{diag}[r_n, ..., r_n^N]$. Based on the above expression and $|\beta_{n,p}^R| = |\beta_{n,p+1}^R| = r_n$ [3–5], the right eigenstate $|\beta_n^R\rangle$ becomes

$$|\beta_n^R\rangle = U_{r_n}|k_n^R\rangle, \tag{88}$$

where $\langle r|U_{r_n}|r'\rangle = r_n^r\delta_{r,r'}$ and $|k_n^R\rangle = c_{n,p}|k_{n,p}^R\rangle + c_{n,p+1}|k_{n,p+1}^R\rangle$ is the superposition of two conventional Bloch waves.

### 2. Left eigenstates

For the left eigenstates, the non-Bloch Hamiltonian of $H_{OBC}^{\dagger}$ is $\mathcal{H}'(\beta) = [\mathcal{H}(1/\beta^*)]^{\dagger}$. To show this, we can expand the Hamiltonian in the form of creation and annihilation operators:

$$H_{OBC} = \sum_{i,j} \sum_{\mu,\nu} t_{ij}^{\mu\nu} \hat{c}_{i\mu}^{\dagger} \hat{c}_{j\nu} \xrightarrow{PBC} \sum_{k} \sum_{\mu,\nu} \sum_{l_{\mu\nu}} t_{l_{\mu\nu}}^{\mu\nu} e^{ikl_{\mu\nu}} \hat{c}_{k\mu}^{\dagger} \hat{c}_{k\nu} \to [\mathcal{H}(\beta)]_{\mu\nu} = \sum_{l_{\mu\nu}} t_{l_{\mu\nu}}^{\mu\nu} \beta^{l_{\mu\nu}} \tag{89}$$

consequently, we obtain

$$\begin{aligned}
H_{OBC}^{\dagger} = \sum_{i,j} \sum_{\mu,\nu} (t_{ji}^{\nu\mu})^* \hat{c}_{i\mu}^{\dagger} \hat{c}_{j\nu} \xrightarrow{PBC} \sum_{k} \sum_{\mu,\nu} \sum_{l_{\nu\mu}} (t_{l_{\nu\mu}}^{\nu\mu})^* e^{-ikl_{\nu\mu}} \hat{c}_{k\mu}^{\dagger} \hat{c}_{k\nu} &\to [\mathcal{H}'(\beta)]_{\mu\nu} = \sum_{l_{\nu\mu}} (t_{l_{\nu\mu}}^{\nu\mu})^* \beta^{-l_{\nu\mu}} \\
&\to [\mathcal{H}'(\beta)]_{\nu\mu} = \sum_{l_{\mu\nu}} (t_{l_{\mu\nu}}^{\mu\nu})^* (1/\beta)^{l_{\mu\nu}} \\
&\to [\mathcal{H}'(\beta)]_{\nu\mu} = \left[ \sum_{l_{\mu\nu}} t_{l_{\mu\nu}}^{\mu\nu} (1/\beta^*)^{l_{\mu\nu}} \right]^* \\
&\to [\mathcal{H}'(\beta)]_{\nu\mu} = [\mathcal{H}(1/\beta^*)_{\mu\nu}]^*.
\end{aligned} \tag{90}$$

This implies $\mathcal{H}'(\beta) = [\mathcal{H}(1/\beta^*)]^{\dagger}$. Therefore the corresponding characteristic equation determined by the $\mathcal{H}'(\beta)$ is

$$f'(\beta, E) = \det[E - \mathcal{H}'(\beta)] = \det[E - [\mathcal{H}(1/\beta^*)]^{\dagger}] = f(1/\beta^*, E^*) = 0. \tag{91}$$

Now, according to the GBZ condition, we know the left eigenstate $|\psi_n^L\rangle$ (whose eigenvalue is $E_n^*$) is a superposition of the following two generalized Bloch waves,

$$|\beta_n^L\rangle := |\psi_n^L\rangle = d_{n,p'}|\beta_{n,p'}^L\rangle + d_{n,p'+1}|\beta_{n,p'+1}^L\rangle, \tag{92}$$

where $\beta_{n,\mu}^L$ is the $\mu$th largest root (ordered by the absolute value) of $f'(\beta, E_n^*) = 0$ and $p'$ is the order of the pole of $f'(\beta, E_n^*) = 0$. According to Eq. 91, i.e., $f'(\beta, E_n^*) = f(1/\beta^*, E_n) = 0$, we have

$$\beta_{n,p'}^L = (1/\beta_{n,p}^R)^* = \frac{e^{ik_{n,p}^R}}{r_n}, \qquad \beta_{n,p'+1}^L = (1/\beta_{n,p+1}^R)^* = \frac{e^{ik_{n,p+1}^R}}{r_n}, \tag{93}$$

which results

$$|\beta_n^L\rangle = U_{1/r_n}|k_n^L\rangle, \tag{94}$$

where $\langle r|U_{1/r_n}|r'\rangle = r_n^{-r}\delta_{r,r'}$ and $|k_n^L\rangle = d_{n,p'}|k_{n,p}^R\rangle + d_{n,p'+1}|k_{n,p+1}^R\rangle$ is the superposition of two conventional Bloch waves. According to Eq. 85, the normalization condition requires

$$\langle\psi_n^L|\psi_n^R\rangle = \langle\beta_n^L|\beta_n^R\rangle = \langle k_n^L|k_n^R\rangle = 1. \tag{95}$$

### 3. Local density of states

Putting Eq. 88 and Eq. 94 into Eq. 84, one can obtain,

$$\begin{aligned}
\nu_i(\omega) &= -\frac{1}{\pi} \text{Im} \left[ \sum_n \frac{\langle i|\psi_n^R\rangle\langle\psi_n^L|i\rangle}{\omega - E_n + i\eta} \right] \\
&= -\frac{1}{\pi} \text{Im} \left[ \sum_n \frac{\langle i|\beta_n^R\rangle\langle\beta_n^L|i\rangle}{\omega - E_n + i\eta} \right] \\
&= -\frac{1}{\pi} \text{Im} \left[ \sum_n \frac{\langle i|k_n^R\rangle\langle k_n^L|i\rangle}{\omega - E_n + i\eta} \right].
\end{aligned} \tag{96}$$

The above formula shows that the local DoS does not depend on the localization length of skin modes, namely, $r_n$. In Fig. 6, we show the comparisons between the local DoS (at different sites, i.e., $\nu_1(\omega)$, $\nu_{30}(\omega)$, $\nu_{70}(\omega)$, $\nu_{100}(\omega)$) with open/periodic boundary conditions. The model Hamiltonian is $\mathcal{H}_{\text{spinless}}(k) - i\gamma$, where the parameters are chosen as $t_1 = 2, t_2 = \mu = \lambda = 1$ and $N = 100$. The gray/black lines represent the local DoS with periodic/open boundaries, respectively. One can notice that $\nu_1(\omega)$ and $\nu_{100}(\omega)$ have the similar behavior, although the system has skin modes localized at site $i = 1$.

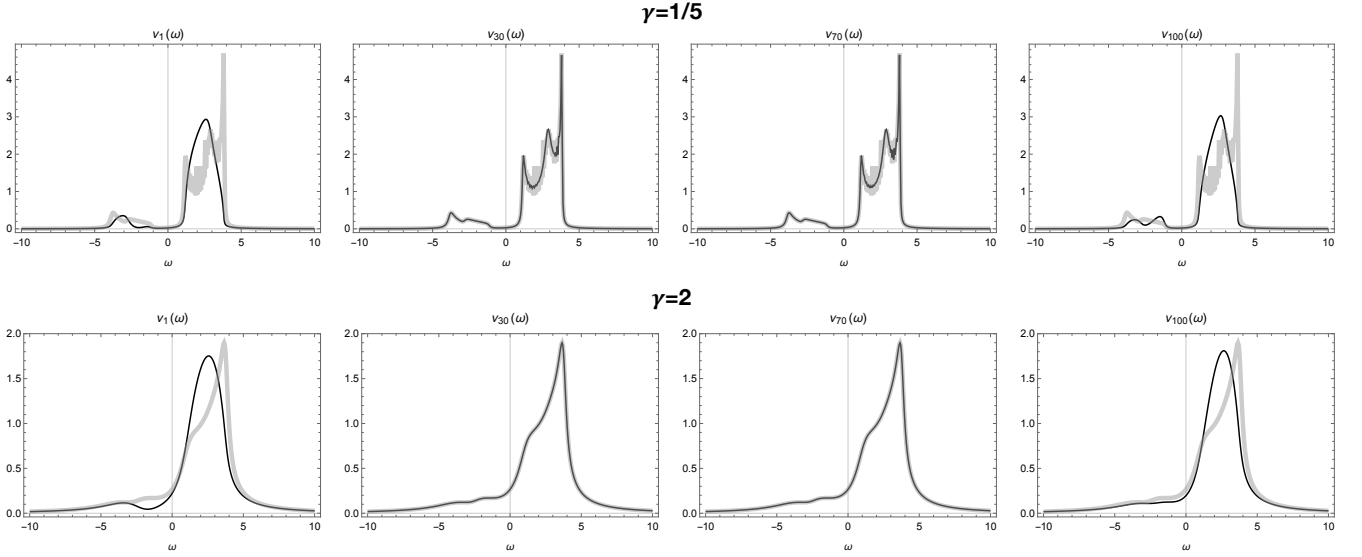

FIG. 6. The local DoS of the spinless model $\mathcal{H}_{\text{spinless}}(k) - i\gamma$ with $t_1 = 2, t_2 = \mu = \lambda = 1, N = 100$ and different values of $\gamma$. The gray/black lines represent local DoS with periodic/open boundaries.

### C. The derivation of chiral tunneling effect

Putting Eq. 88 and Eq. 94 into

$$\langle f|U(t)|i\rangle = \langle f|e^{-i\mathcal{H}t}|i\rangle = \sum_n e^{-iE_n t}\langle f|\psi_n^R\rangle\langle\psi_n^L|i\rangle, \tag{97}$$

we can obtain

$$\begin{aligned}\langle f|U(t)|i\rangle &= \sum_n e^{-iE_n t}\langle f|\beta_n^R\rangle\langle\beta_n^L|i\rangle \\ &= \sum_n r_n^{f-i} e^{-iE_n t}\langle f|k_n^R\rangle\langle k_n^L|i\rangle,\end{aligned} \tag{98}$$

where $r_n = \left|\beta_{n,p}^R\right| = \left|\beta_{n,p+1}^R\right|$ represents the localization length of skin modes. Consequently, we have $P_{N\leftarrow 1}(t) = |\langle N|U(t)|1\rangle|^2 \propto r_n^{N-1}$ and $P_{1\leftarrow N}(t) = |\langle 1|U(t)|N\rangle|^2 \propto r_n^{1-N}$, that is, the strength of the asymmetric tunneling exponentially depends on the localization length of skin modes.

### D. Bounded tunneling probability

When $r_n < 1$, it seems that the term $r_n^{1-N}$ will tend to infinity in the limit of $N \to \infty$. Now we show that although there is a seemingly divergent term $r_n^{f-i}$, the tunneling probability

$$P_{f\leftarrow i}(t) = |\langle f|U(t)|i\rangle|^2 \tag{99}$$

is bounded to less than 1, when the system is purely dissipative.

First, we notice that

$$\langle i|U^\dagger(t)U(t)|i\rangle = \sum_f |G_{f\leftarrow i}(t)|^2 = \sum_f P_{f\leftarrow i}(t). \tag{100}$$

If we take $t = 0$ in the above equation, we can obtain $\langle i|U^\dagger(t=0)U(t=0)|i\rangle = \sum_f |G_{f\leftarrow i}(t=0)|^2 = 1$, which means $P_{f\leftarrow i}(t=0) = |G_{f\leftarrow i}(t=0)|^2 \leq 1$. On the other hand, we have

$$\langle i|U^\dagger(t)U(t)|i\rangle = \sum_{mn} \langle i|\psi_m^L\rangle\langle\psi_m^R|\psi_n^R\rangle\langle\psi_n^L|i\rangle e^{i(E_m^* - E_n)t} \equiv \sum_{mn} T_{mn}, \tag{101}$$

where

$$T_{mn} = \langle i|\psi_m^L\rangle\langle\psi_m^R|\psi_n^R\rangle\langle\psi_n^L|i\rangle e^{i(E_m^* - E_n)t} \tag{102}$$

and $|\psi_m^{R/L}\rangle$ represent the right/left eigenstates, which can be skin modes or non-skin modes.

If we take $m = n$, we will have

$$T_{nn} = \langle i|\psi_n^L\rangle\langle\psi_n^R|\psi_n^R\rangle\langle\psi_n^L|i\rangle e^{i(E_n^* - E_n)t}. \tag{103}$$

Since the system is dissipative, the imaginary part of the eigenvalue must be negative. Thus without loss of generality, we can write $E_n = a_n - ib_n$ with $b_n > 0$ (dissipation), then

$$T_{nn} = \langle i|\psi_n^L\rangle\langle\psi_n^R|\psi_n^R\rangle\langle\psi_n^L|i\rangle e^{-2b_n t}, \tag{104}$$

which will decay with time.

On the other hand, if $m \neq n$, then

$$T_{mn} + T_{nm} = c_{mn} e^{i(E_m^* - E_n)t} + c_{mn}^* e^{i(E_n^* - E_m)t} \propto e^{-b_{mn}t}, \tag{105}$$

where

$$b_{mn} = \text{Im}[E_m^* - E_n] = b_m + b_n, \tag{106}$$

which will also decay with time.

Therefore, we finally obtain that $\langle i|U^\dagger(t)U(t)|i\rangle$ will decay with time, which gives the upper bounds of $P_{f\leftarrow i}(t)$

$$P_{f\leftarrow i}(t) = \sum_f P_{f\leftarrow i}(t) = \langle i|U^\dagger(t)U(t)|i\rangle \leq \langle i|U^\dagger(0)U(0)|i\rangle = 1. \tag{107}$$

## VII. MATHEMATICA CODE

In the following pages, we provide a *Mathematica* code to calculate the auxiliary GBZ [5] of the spinful model Eq. 5 proposed in the main text.

---

### The Bloch Hamiltonian

```mathematica
In[ ]:= Hk[k_, t1_, t2_, λI_, λR_, μ_, γ_] := (t1 + t2 Cos[k]) ArrayFlatten[TensorProduct[{{0, 1}, {1, 0}}, {{1, 0}, {0, 1}}]] +
    t2 Sin[k] ArrayFlatten[TensorProduct[{{0, -I}, {I, 0}}, {{1, 0}, {0, 1}}]] +
    μ ArrayFlatten[TensorProduct[{{1, 0}, {0, -1}}, {{1, 0}, {0, 1}}]] +
    λI Sin[k] ArrayFlatten[TensorProduct[{{1, 0}, {0, -1}}, {{1, 0}, {0, -1}}]] -
    λR (1 / 2 ArrayFlatten[TensorProduct[{{0, -I}, {I, 0}}, {{0, 1}, {1, 0}}]] -
    3 ^ (1 / 2) / 2 ArrayFlatten[TensorProduct[{{0, -I}, {I, 0}}, {{0, -I}, {I, 0}}]]) +
    I γ ArrayFlatten[TensorProduct[{{1, 0}, {0, -1}}, {{1, 0}, {0, 1}}]]];
    Hk[k, t1, t2, λI, λR, μ, γ] // MatrixForm
```

Out[ ]//MatrixForm=

$$\begin{pmatrix} i\,\gamma + \mu + \lambda I\,\text{Sin}[k] & 0 & t1 + t2\,\text{Cos}[k] - i\,t2\,\text{Sin}[k] & -\left(\frac{i}{2} + \frac{\sqrt{3}}{2}\right)\lambda R \\ 0 & i\,\gamma + \mu - \lambda I\,\text{Sin}[k] & -\left(\frac{i}{2} - \frac{\sqrt{3}}{2}\right)\lambda R & t1 + t2\,\text{Cos}[k] - i\,t2\,\text{Sin}[k] \\ t1 + t2\,\text{Cos}[k] + i\,t2\,\text{Sin}[k] & -\left(\frac{i}{2} - \frac{\sqrt{3}}{2}\right)\lambda R & -i\,\gamma - \mu - \lambda I\,\text{Sin}[k] & 0 \\ -\left(\frac{i}{2} + \frac{\sqrt{3}}{2}\right)\lambda R & t1 + t2\,\text{Cos}[k] + i\,t2\,\text{Sin}[k] & 0 & -i\,\gamma - \mu + \lambda I\,\text{Sin}[k] \end{pmatrix}$$

### The real space Hamiltonian

```mathematica
In[ ]:= Hr[t1_, t2_, λI_, λR_, μ_, γ_, n_] :=
    SparseArray[Band[{1, 1}, {n, n}] → {Coefficient[TrigToExp[Hk[k, t1, t2, λI, λR, μ, γ]], Exp[I k], 0]}, {n, n}] +
    SparseArray[Band[{5, 1}, {n, n}] → {Coefficient[TrigToExp[Hk[k, t1, t2, λI, λR, μ, γ]], Exp[I k], -1]}, {n, n}] +
    SparseArray[Band[{1, 5}, {n, n}] → {Coefficient[TrigToExp[Hk[k, t1, t2, λI, λR, μ, γ]], Exp[I k], 1]}, {n, n}];
```

### Periodic/open boundary spectra

```mathematica
In[ ]:= PeriodicSpectra = Flatten[Table[Eigenvalues[N[Hk[k, 2, 1, 2, 1, 1, 1]]], {k, -3.14, 3.14, 3.14 / 25}], 1];
    (*N=100/4≈25*) OpenSpectra = Eigenvalues[N[Hr[2, 1, 2, 1, 1, 1, 100], 50]];
    Spectra = ListPlot[{(Tooltip[{Re[#1], Im[#1]}] &) /@ PeriodicSpectra, (Tooltip[{Re[#1], Im[#1]}] &) /@ OpenSpectra},
      PlotTheme → "Scientific", AspectRatio → 1, PlotRange → All,
      PlotStyle → {{GrayLevel[0.7], PointSize[0.025]}, {GrayLevel[0.2], PointSize[0.02]}, {GrayLevel[0.1], PointSize[0.01]}},
      ImageSize → 350, FrameLabel → {{HoldForm["Im[E]"], None}, {HoldForm["Re[E]"], None}},
      PlotLabel → HoldForm[Spectra], LabelStyle → {GrayLevel[0.4], 15}]
```

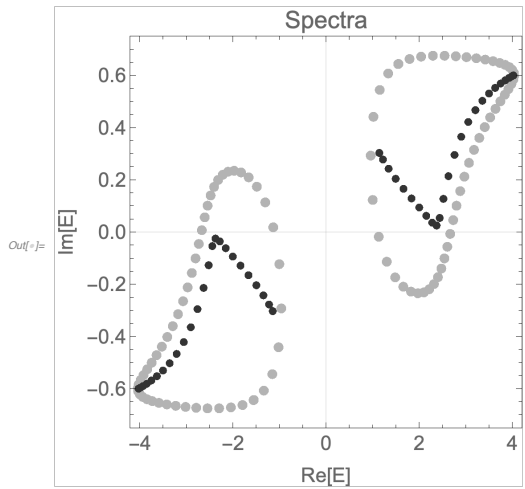

### Non-Bloch Hamiltonian and its characteristic equation

```mathematica
In[ ]:= Hb[β_, t1_, t2_, λI_, λR_, μ_, γ_] := TrigToExp[Hk[k, t1, t2, λI, λR, μ, γ]] /. {Exp[I k] → β, Exp[-I k] → 1 / β}
    Hb[β, t1, t2, λI, λR, μ, γ] // MatrixForm
    F[β_, Ene_, t1_, t2_, λI_, λR_, μ_, γ_] := Factor[Det[Ene IdentityMatrix[4] - Hb[β, t1, t2, λI, λR, μ, γ]]];
```

Out[ ]//MatrixForm=

$$\begin{pmatrix} i\,\gamma + \frac{i\,\lambda I}{2\,\beta} - \frac{i\,\beta\,\lambda I}{2} + \mu & 0 & t1 + \frac{t2}{\beta} & \frac{i\,\lambda R}{2} - \frac{\sqrt{3}\,\lambda R}{2} \\ 0 & i\,\gamma - \frac{i\,\lambda I}{2\,\beta} + \frac{i\,\beta\,\lambda I}{2} + \mu & \frac{i\,\lambda R}{2} + \frac{\sqrt{3}}{2}\,\lambda R & t1 + \frac{t2}{\beta} \\ t1 + t2\,\beta & -\frac{i\,\lambda R}{2} + \frac{\sqrt{3}\,\lambda R}{2} & -i\,\gamma - \frac{i\,\lambda I}{2\,\beta} + \frac{i\,\beta\,\lambda I}{2} - \mu & 0 \\ -\frac{i\,\lambda R}{2} - \frac{\sqrt{3}\,\lambda R}{2} & t1 + t2\,\beta & 0 & -i\,\gamma + \frac{i\,\lambda I}{2\,\beta} - \frac{i\,\beta\,\lambda I}{2} - \mu \end{pmatrix}$$

### Auxiliary GBZ (arXiv:1912.05499)

### 1. Eliminating E

FIG. 7. Mathematica code

```
In[*]:= Res[β1_, β2_, t_, t1_, t2_, λI_, λR_, μ_, γ_] :=
         Resultant[F[β1+I β2, Ene, t1, t2, λI, λR, μ, γ], F[(β1+I β2) ((1-t^2+2 I t) / (1+t^2)), Ene, t1, t2, λI, λR, μ, γ], Ene];
       g0 = FactorList[Factor[Res[β1, β2, t, 2, 1, 2, 1, 1, 1] ((-i+t)^16 (i+t)^16 (β1+i β2)^16) / (256 t^4)]]
```

```
Out[*]= {{1, 1}, {i + t - i β1^2 + t β1^2 + 2 β1 β2 + 2 i t β1 β2 + i β2^2 - t β2^2, 4},
       {4 t^2 - 8 i t^3 - 4 t^4 - 8 t^2 β1 + 8 i t^3 β1 - 8 t^4 β1 + 8 i t^5 β1 - (5 + 8 i) β1^2 + (2 - 16 i) t^2 β1^2 + (11 - 8 i) t^4 β1^2 +
         4 t^6 β1^2 - 8 t^2 β1^3 - 8 i t^3 β1^3 - 8 t^4 β1^3 - 8 i t^5 β1^3 + 4 t^2 β1^4 - 4 t^4 β1^4 - 8 i t^2 β2 - 8 t^3 β2 - 8 i t^4 β2 - 8 t^5 β2 +
         (16 - 10 i) β1 β2 + (32 + 4 i) t^2 β1 β2 + (16 + 22 i) t^4 β1 β2 + 8 i t^6 β1 β2 - 24 i t^2 β1^2 β2 + 24 t^3 β1^2 β2 - 24 i t^4 β1^2 β2 +
         24 t^5 β1^2 β2 + 16 i t^2 β1^3 β2 - 32 t^3 β1^3 β2 - 16 i t^4 β1^3 β2 + (5 + 8 i) β2^2 - (2 - 16 i) t^2 β2^2 - (11 - 8 i) t^4 β2^2 - 4 t^6 β2^2 +
         24 t^2 β1 β2^2 + 24 i t^3 β1 β2^2 - 24 t^4 β1 β2^2 + 24 i t^5 β1 β2^2 - 48 i t^3 β1^2 β2^2 + 24 t^4 β1^2 β2^2 - 48 i t^2 β2^3 +
         8 t^3 β2^3 + 8 i t^4 β2^3 - 8 t^5 β2^3 - 16 i t^2 β1 β2^3 + 32 t^3 β1 β2^3 + 16 i t^4 β1 β2^3 + 4 t^2 β2^4 + 8 i t^3 β2^4 - 4 t^4 β2^4, 2},
```

[large resultant factor output continues]

```
{-4 + 16 i t + 24 t^2 - 16 i t^3 - 4 t^4 + 8 β1 - 24 i t β1 - 16 t^2 β1 - 16 i t^3 β1 - 24 t^4 β1 + 8 i t^5 β1 - (9 + 8 i) β1^2 - (16 - 18 i) t β1^2 -
 (9 + 8 i) t^2 β1^2 - (32 - 36 i) t^3 β1^2 + (9 + 8 i) t^4 β1^2 - (16 - 18 i) t^5 β1^2 + (9 + 8 i) t^6 β1^2 - 8 β1^3 + 8 i t β1^3 - 16 t^2 β1^3 + 16 i t^3 β1^3 -
 8 t^4 β1^3 + 8 i t^5 β1^3 + (6 - 16 i) β1^4 + (10 - 48 i) t^2 β1^4 + (2 - 48 i) t^4 β1^4 - (2 + 16 i) t^6 β1^4 - 8 β1^5 - 8 i t β1^5 - 16 t^2 β1^5 - 16 i t^3 β1^5 -
 8 t^4 β1^5 - 8 i t^5 β1^5 - (9 + 8 i) β1^6 + (9 + 8 i) t^6 β1^6 + (16 - 18 i) t β1^6 + (9 + 8 i) t^2 β1^6 + (16 - 18 i) t^5 β1^6 +
 (9 + 8 i) t^6 β1^6 + 8 β1^7 - 24 i t β1^7 - 16 t^2 β1^7 + 16 i t^3 β1^7 - 24 t^4 β1^7 - 8 i t^5 β1^7 - 4 β1^8 - 16 i t β1^8 + 24 t^2 β1^8 + 16 i t^3 β1^8 -
 4 t^4 β1^8 + 8 i t^5 β2 + 24 t β2 - 16 i t^2 β2 + 16 t^3 β2 - 8 t^5 β2^3 β2 - (16 - 32 i) t β1 β2 - (16 - 18 i) t^2 β1 β2 -
 (72 + 64 i) t^3 β1 β2 - (16 - 18 i) t^4 β1 β2 - (36 - 32 i) t^5 β1 β2 - (16 - 18 i) t^6 β1 β2 - 24 i t^2 β1^2 β2 - 24 t^3 β1^2 β2 - 48 i t^2 β1^2 β2 -
 48 t^3 β1^2 β2 - 24 i t^4 β1^2 β2 - 24 t^5 β1^2 β2 + (64 - 8 i) β1^3 β2 + (192 + 40 i) t^2 β1^3 β2 + (64 - 8 i) t^4 β1^3 β2 -
 40 i β1^4 β2 + 40 t β1^4 β2 - 80 i t^2 β1^4 β2 + 80 t^3 β1^4 β2 - 40 i t^4 β1^4 β2 + 40 t^5 β1^4 β2 + (48 - 54 i) β1^5 β2 + (108 + 96 i) t β1^5 β2 +
 (48 - 54 i) t^2 β1^5 β2 + (216 + 192 i) t^3 β1^5 β2 - (48 - 54 i) t^4 β1^5 β2 + (108 + 96 i) t^5 β1^5 β2 - (48 - 54 i) t^6 β1^5 β2 + 56 i β1^6 β2 -
 168 t β1^6 β2 - 112 i t^2 β1^6 β2 - 112 t^3 β1^6 β2 - 168 i t^4 β1^6 β2 + 56 t^5 β1^6 β2 - 32 i β1^7 β2 + 128 t β1^7 β2 + 192 i t^2 β1^7 β2 - 128 t^3 β1^7 β2 -
 32 i t^4 β1^7 β2 + (9 + 8 i) β2^2 + (16 - 18 i) t β2^2 - (9 + 8 i) t^2 β2^2 + (32 - 36 i) t^3 β2^2 - (9 + 8 i) t^4 β2^2 + (16 - 18 i) t^5 β2^2 - (9 + 8 i) t^6 β2^2 +
 24 β1 β2^2 - 24 i t β1 β2^2 + 48 t^2 β1 β2^2 - 48 i t^3 β1 β2^2 + 24 t^4 β1 β2^2 + 24 i t^5 β1 β2^2 - (36 - 96 i) t^2 β1^2 β2^2 - (60 - 288 i) t^2 β1^2 β2^2 -
 (12 - 288 i) t^4 β1^2 β2^2 + (12 - 96 i) t^6 β1^2 β2^2 + 80 i t^3 β1^3 β2^2 + 160 t^2 β1^3 β2^2 + 160 i t^3 β1^3 β2^2 + 80 t^4 β1^3 β2^2 + 80 i t^5 β1^3 β2^2 +
 (135 + 120 i) β1^4 β2^2 - (240 - 270 i) t β1^4 β2^2 + (135 + 120 i) t^2 β1^4 β2^2 - (480 - 540 i) t^3 β1^4 β2^2 - (135 + 120 i) t^4 β1^4 β2^2 -
 (240 - 270 i) t^5 β1^4 β2^2 - (135 + 120 i) t^6 β1^4 β2^2 - 168 i β1^5 β2^2 - 504 t β1^5 β2^2 + 336 t^2 β1^5 β2^2 - 336 i t^3 β1^5 β2^2 - 504 t^4 β1^5 β2^2 +
 168 i t^5 β1^5 β2^2 + 112 β1^6 β2^2 + 448 i t β1^6 β2^2 - 672 t^2 β1^6 β2^2 - 448 i t^3 β1^6 β2^2 + 112 t^4 β1^6 β2^2 + 8 i β2^3 + 8 t β2^3 + 16 i t^2 β2^3 +
 16 t^3 β2^3 + 8 i t^4 β2^3 + 8 t^5 β2^3 - (64 + 24 i) β1 β2^3 - (192 + 40 i) t^2 β1 β2^3 - (64 - 8 i) t^4 β1 β2^3 -
 80 t β1^2 β2^3 + 160 i t^2 β1^2 β2^3 - 160 t^3 β1^2 β2^3 + 80 i t^4 β1^2 β2^3 - 80 t^5 β1^2 β2^3 - (160 - 180 i) β1^3 β2^3 - (360 + 320 i) t β1^3 β2^3 -
 (160 - 180 i) t^2 β1^3 β2^3 - (720 + 640 i) t^3 β1^3 β2^3 - (160 - 180 i) t^4 β1^3 β2^3 - (360 + 320 i) t^5 β1^3 β2^3 + (160 - 180 i) t^6 β1^3 β2^3 -
 280 i β1^4 β2^3 + 840 t β1^4 β2^3 + 560 i t^2 β1^4 β2^3 + 560 t^3 β1^4 β2^3 + 840 i t^4 β1^4 β2^3 - 280 t^5 β1^4 β2^3 - 224 i β1^5 β2^3 - 896 t β1^5 β2^3 -
 1344 i t^2 β1^5 β2^3 + 896 t^3 β1^5 β2^3 + 224 i t^4 β1^5 β2^3 + (6 - 16 i) β2^4 + (10 - 48 i) t^2 β2^4 - (2 + 16 i) t^6 β2^4 -
 40 β1 β2^4 - 40 i t β1 β2^4 - 80 t^2 β1 β2^4 - 80 i t^3 β1 β2^4 - 40 t^4 β1 β2^4 - 40 i t^5 β1 β2^4 - (135 + 120 i) β1^2 β2^4 + (240 - 270 i) t β1^2 β2^4 -
 (135 + 120 i) t^2 β1^2 β2^4 + (480 - 540 i) t^3 β1^2 β2^4 + (135 + 120 i) t^4 β1^2 β2^4 + (240 - 270 i) t^5 β1^2 β2^4 + (135 + 120 i) t^6 β1^2 β2^4 +
 280 β1^3 β2^4 + 840 i t β1^3 β2^4 - 560 t^2 β1^3 β2^4 + 560 i t^3 β1^3 β2^4 + 840 t^4 β1^3 β2^4 - 280 β1^4 β2^4 - 1120 i t β1^4 β2^4 +
 1680 t^2 β1^4 β2^4 + 1120 i t^3 β1^4 β2^4 - 280 t^4 β1^4 β2^4 - 8 i β2^5 + 8 t β2^5 - 16 i t^2 β2^5 + 16 t^3 β2^5 - 8 i t^4 β2^5 + 8 t^5 β2^5 - (48 - 54 i) β1 β2^5 +
 (108 + 96 i) t β1 β2^5 + (48 - 54 i) t^2 β1 β2^5 + (216 + 192 i) t^3 β1 β2^5 - (48 - 54 i) t^4 β1 β2^5 + (108 + 96 i) t^5 β1 β2^5 - (48 - 54 i) t^6 β1 β2^5 +
 168 i β1^2 β2^5 - 504 t β1^2 β2^5 - 336 i t^2 β1^2 β2^5 - 336 t^3 β1^2 β2^5 - 504 i t^4 β1^2 β2^5 + 168 t^5 β1^2 β2^5 - 224 i β1^3 β2^5 + 896 t β1^3 β2^5 +
 1344 i t^2 β1^3 β2^5 - 896 t^3 β1^3 β2^5 - 224 i t^4 β1^3 β2^5 + (9 + 8 i) β2^6 - (16 - 18 i) t β2^6 + (9 + 8 i) t^2 β2^6 - (32 - 36 i) t^3 β2^6 - (9 + 8 i) t^4 β2^6 -
 (16 - 18 i) t^5 β2^6 - (9 + 8 i) t^6 β2^6 - 56 β1 β2^6 - 168 i t β1 β2^6 + 112 t^2 β1 β2^6 - 112 i t^3 β1 β2^6 + 168 t^4 β1 β2^6 + 56 i t^5 β1 β2^6 + 112 β1^2 β2^6 +
 448 i t β1^2 β2^6 - 672 t^2 β1^2 β2^6 - 448 i t^3 β1^2 β2^6 + 112 t^4 β1^2 β2^6 + 8 i β2^7 + 24 t β2^7 + 16 i t^2 β2^7 + 16 t^3 β2^7 + 24 i t^4 β2^7 - 8 t^5 β2^7 +
 32 i β1 β2^7 - 128 t β1 β2^7 - 192 i t^2 β1 β2^7 + 128 t^3 β1 β2^7 + 32 i t^4 β1 β2^7 - 4 β2^8 - 16 i t β2^8 + 24 t^2 β2^8 + 16 i t^3 β2^8 - 4 t^4 β2^8, 2}}
```

## 2. Eliminating t for each factor

```
In[*]:= g1a = ComplexExpand[Re[g0[[2]][[1]]]];
       g1b = ComplexExpand[Im[g0[[2]][[1]]]];
       aGBZ1 = Resultant[g1a, g1b, t];
```

```
In[*]:= g2a = ComplexExpand[Re[g0[[3]][[1]]]];
       g2b = ComplexExpand[Im[g0[[3]][[1]]]];
       aGBZ2 = Factor[Resultant[g2a, g2b, t]];
```

```
In[*]:= g3a = ComplexExpand[Re[g0[[4]][[1]]]];
       g3b = ComplexExpand[Im[g0[[4]][[1]]]];
       aGBZ3 = Factor[Resultant[g3a, g3b, t]];
```

FIG. 8. Mathematica code

```
In[ ]:= aGBZ = ContourPlot[{aGBZ1 ⩵ 0, aGBZ2 ⩵ 0, aGBZ3 ⩵ 0}, {β1, -1.5, 1.5}, {β2, -1.5, 1.5},
      PlotTheme → "Scientific", ImageSize → 350, ContourStyle → {{RGBColor[1, 0, 0, 0.5], Thickness[0.01]},
        {RGBColor[1, 0, 0, 0.5], Thickness[0.01]}, {RGBColor[1, 0, 0, 0.5], Thickness[0.01]}},
      FrameLabel → {{HoldForm["Im[β]"], None}, {HoldForm["Re[β]"], None}}, PlotLabel → HoldForm[aGBZ],
      LabelStyle → {GrayLevel[0.4], 15}]
```

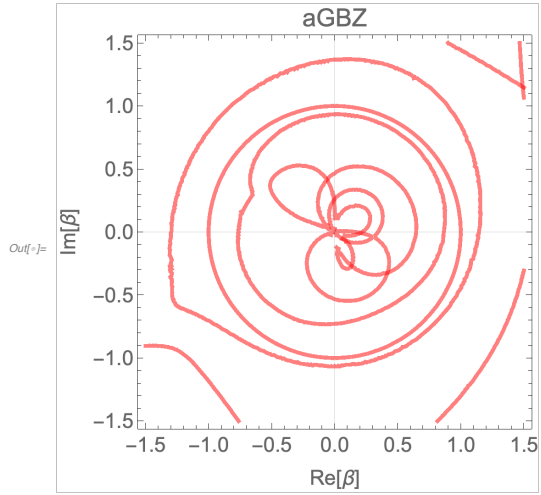

## GBZ

```
In[ ]:= GBZ1 = Table[Sort[z /. NSolve[F[z, OpenSpectra[[i]]], 2, 1, 2, 1, 1, 1], z], Abs[#1] < Abs[#2] &][[3]], {i, 1, 100}];
      GBZ2 = Table[Sort[z /. NSolve[F[z, OpenSpectra[[i]]], 2, 1, 2, 1, 1, 1], z], Abs[#1] < Abs[#2] &][[4]], {i, 1, 100}];
      GBZ3 = Table[Sort[z /. NSolve[F[z, OpenSpectra[[i]]], 2, 1, 2, 1, 1, 1], z], Abs[#1] < Abs[#2] &][[5]], {i, 1, 100}];
      GBZ4 = Table[Sort[z /. NSolve[F[z, OpenSpectra[[i]]], 2, 1, 2, 1, 1, 1], z], Abs[#1] < Abs[#2] &][[6]], {i, 1, 100}];
```

```
In[ ]:= Show[{aGBZ, Graphics[{Table[{Black, Opacity[0.8], PointSize[0.02], Point[{Re[GBZ1[[i]]], Im[GBZ1[[i]]]}]}, {i, 1, 100}],
        Table[{Black, Opacity[0.8], PointSize[0.02], Point[{Re[GBZ2[[i]]], Im[GBZ2[[i]]]}]}, {i, 1, 100}],
        Table[{Black, Opacity[0.8], PointSize[0.02], Point[{Re[GBZ3[[i]]], Im[GBZ3[[i]]]}]}, {i, 1, 100}],
        Table[{Black, Opacity[0.8], PointSize[0.02], Point[{Re[GBZ4[[i]]], Im[GBZ4[[i]]]}]}, {i, 1, 100}]}]}]]
```

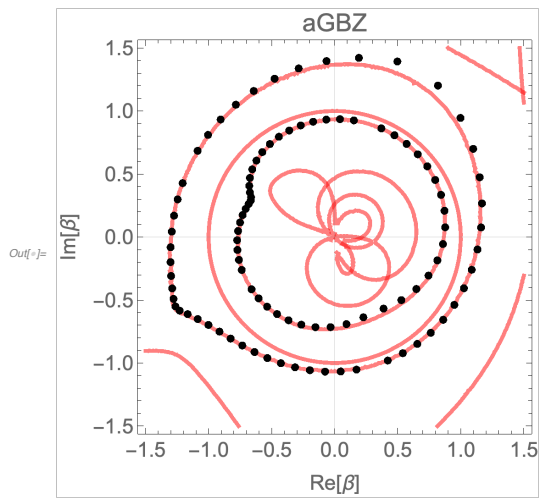

FIG. 9. Mathematica code


	
\end{document}